\documentstyle[12pt]{article}

\newcommand{\be}{\begin{equation}}
\newcommand{\ee}{\end{equation}}

\newcommand{\bea}{\begin{eqnarray}}
\newcommand{\eea}{\end{eqnarray}}

\newcommand{\p}{\partial}

\newcommand{\nn}{\nonumber \\}
\newcommand{\f}{\frac}
\newcommand{\w}{\wedge}

\begin{document}
\thispagestyle{empty}

\begin{flushright}
{\bf arXiv: 1006.2444}
\end{flushright}
\begin{center} \noindent \Large \bf
Quantum phase transition in a Dp-Dq system 

\end{center}

\bigskip\bigskip\bigskip
\vskip 0.5cm
\begin{center}
{ \normalsize \bf   Shesansu Sekhar Pal}

\vskip 0.5cm

\vskip 0.5 cm
Center for Quantum Spacetime, \\
Sogang University, 121-742 Seoul, South Korea\\
\vskip 0.5 cm
\sf shesansu${\frame{\shortstack{AT}}}$gmail.com
\end{center}
\centerline{\bf \small Abstract}

Using  the top-down approach, we study  intersecting Dp-Dq brane configuration in string theory and find  examples, where there can be a quantum phase transition at zero temperature induced by the violation of the Breitenlohner-Freedman (BF) bound at IR, which is done essentially by a combination of charge density and magnetic fields. In particular, there exists a Berezinskii-Kosterlitz-Thouless (BKT) type of transition for D3-D5   and D5-D5 systems. The study of the BKT type of transition is initiated by Jensen et al. [Phys. Rev. Lett. {\bf 105}, 041601 (2010) ] for a D3-D5 system with nonzero charge density and a magnetic field. Here, we show that one can have the BKT transition for a D3-D5 system 
even in the absence of charge density but requires multiple magnetic fields. In this case the field theory lives in 2+1 dimensions, whereas for the D5-D5 type,  the transition requires the presence of both the charge density and magnetic fields and the dual field theory lives on a 3+1 dimensional spacetime. We also study the D3-D7 system but it does not show the BKT type of transition.
\newpage
\section{Introduction}

The study of holographic phase transitions are interesting in their own right as these studies might shed some light on the microscopic-understanding of the related systems  at the strongly coupled limit. It is even more important to  find such   examples 
where the dominant energy is the zero point energy of quantum physics that makes  the phase transition  occur \cite{qpt}, i.e. either at zero temperature or at very low temperature. In  recent developments, it is shown for systems that falls under the AdS/CFT correspondence regime \cite{jm}, how such a phase transition is induced when the mass of the scalar field which is dual to a real or complex valued order parameter goes below the BF bound, especially in the presence of external electric and magnetic fields. In particular,  there exists  phase transition for some particular systems for which it is said to be of infinite order and which resemble that of the BKT type \cite{bkt}\footnote{Generically the order of phase transition could be anything for a system showing quantum phase transition, see for example \cite{jkt} and \cite{ilms} for the second order transition, so also in the study of  high temperature holographic superconductor in \cite{ss},\cite{hhh}, \cite{dh},\cite{gsw} and \cite{bmp} and in some other earlier studies e.g. \cite{hkst}.}.

This is achieved in  string theory using the top-down approach by considering  an intersecting brane  configuration of D3 and D5 types \cite{jkst}. These branes are extended in such a way 
that they have got only four number of directions along which they satisfy the ND+DN boundary conditions. The external  electric and magnetic fields are put along the intersecting  directions of this brane configuration and the dual field theory lives on a 2+1 dimension.  However, if we consider a T-dual configuration to this system namely, intersecting D3-D7 branes   with electric and magnetic fields along the intersecting directions, which is a 3+1 dimensional field theory and it does not fall in the  type of theories that shows BKT kind of phase transition. Rather it is suggested that it shows a second order phase transition \cite{jkt}. In the absence of a magnetic field such a configuration was studied in \cite{ko} but at finite temperature.  

It is not {\em a priori} clear why the earlier  brane  configuration showed a BKT type phase transition but not the later type  ? One possibility as suggested in \cite{jkst} is that, perhaps,  the dimension of both the charge density and the magnetic field are the same for the 2+1 dimensional field theory  but not for the 3+1 dimensional field theory. 
It also looks like both the charge density and the magnetic fields are essentially to see the BKT type transition. 
In what follows, we shall see in an example that  this is not the complete story, as we shall see the BKT-type transition exists even for systems with  zero density, for example in D3-D5 brane configuration. 

Originally, the BKT transition \cite{bkt} is found at nonzero temperature with the condensate  behaving as $e^{-\f{c}{\sqrt{T_c-T}}}$, which we shall refer to as the order parameter,  near to critical temperature $T_c$. Here, on the other hand at zero temperature the order parameter behaves as $e^{-\f{c}{\sqrt{\nu_c-\nu}}}$ near to the critical parameter $\nu_c$, where  $\nu$ depends on either a specific combination  of  charge density and magnetic fields or only on magnetic fields. The parameter $\nu_c$ is determined  when the mass of the dual bulk field related to this order parameter saturates the BF bound and $c$ has the structure $c=\pi\sqrt{\f{\nu^2+1}{\nu+\nu_c}}$. For $\nu <\nu_c$, it describes a system without the chiral symmetry i.e. there exists a nontrivial condensate to the operator dual to the massive scalar field $y$, whereas for
 $\nu > \nu_c$, it describes the system with the chiral symmetry which means zero condensate. It is interesting to note that, for $\nu$ close to $\nu_c$, the asymptotic AdS solution is unstable against the perturbation to scalar field, so   it will  drive the system  from a chirally symmetric phase to an asymmetric phase. In terms of gravity solution the BKT type of transition goes from a solution which is zero, i.e. $y=0$,   to a solution where $y\neq 0$; essentially it is the bulk field $y$ that  describes this transition. The chiral symmetry here corresponds to the R-symmetry\footnote{Note that Weyl fermions do not exists in   odd dimensional spacetime. Hence  the chiral symmetry in the strict sense does not exist in 2+1 dimension, but here we mean it as the R symmetry. }. 

The question that we address in this paper is to start constructing different types of intersecting  brane configurations like Dp and Dq branes at zero temperature which are supersymmetric and hence are stable. After turning on the external fields like electric   and several constant magnetic field makes the system nonsupermmetric and hence could be unstable. The easiest way to see it's nonsupersymmetric nature is from eq(\ref{quadratic_action_b1_b2_b3}), where there arises a nontrivial  potential energy term for the scalar field $y$, of course for nonvanishing electric and magnetic fields. So,  it does not obey the no-force condition which is essential to show the supersymmetric nature.

For a specific range of  parameter $\nu <\nu_c$ and for few intersecting brane configurations, 
 there occurs a violation to the BF bound at IR,  which  essentially signals the presence of instability and as a consequence leads to a phase transition. Strictly,  it leads to the
appearance of quantum phase transition because we are at zero temperature. This idea is followed and studied rigorously in all of the phase transitions in holographic superconductivity \cite{ss}, \cite{hhh} and \cite{gsw}, so also in \cite{ilms} at the low temperature regime.

A recent study \cite{jkst} found yet another situation where the above-mentioned philosophy is followed and found the unique exponentially suppressing behavior to the order parameter, which is  suggested as the holographically generated  
BKT type of phase transition. This simply came out of the  D3-D5 brane configuration  with an electric and magnetic field turned on the world volume of the D5 brane, whose  field theory  lives in a 2+1 dimensional spacetime. It so happens that in this example of the D3-D5 brane  system, the dimension of the charge density and the magnetic field are the same. It is also suggested in \cite{jkst} that this very particular behavior of charge density and magnetic field plays a crucial role in showing the BKT type of transition. 

The study of   other intersecting brane systems with external fields, which we shall describe, shows that  the BKT type of transition somehow follows this "thumb rule". Examples that suggest this are  the D3-D5 system with two different magnetic fields  and with or without the electric field and the D5-D5 system with both the electric and magnetic field, which lives in a 3+1 dimensional field theory. A dimensionless combination of these external fields induces the transition to occur. 
Certainly, it is important  to study and find  the appearance  of BKT  type of phase transition in other kinds of brane configurations, which  might give a better understanding of this phenomena. Another important point to note is that, in contrast to the original BKT phase transition \cite{bkt}, these holographic field theories live in more than two dimensional  spaces in Euclidean signature. However,  both of these brane configurations see $AdS_2$ at IR with a parameter $\nu$, which could be the plausible reason to observe such a kind of transition.

For the Dp-Dq system under investigation, we shall assume that the  magnetic fields of constant  strengths $B_1$ and $B_2$ are put on the world volume common to both the branes and another  magnetic field of strength $B_3$ is put along the direction perpendicular to the Dp brane, along with an electric field on the world volume of the Dq brane. In this Dp-Dq system, it is this external electric and magnetic fields that cause the instability, essentially by making the mass of the embedding scalar field to go below the BF bound in the  probe brane approximation.   In this  approximation \cite{kk}, one essentially considers a limit in which the number of probe Dq branes are very small in comparison to the number of Dp branes. So  the energy density associated to the Dp brane dominates over that of the Dq brane and one can safely forget about considering the back reaction of the probe Dq branes on the Dp branes at the leading order. 

In particular, for the D3-D5 brane system, the  scalar field $y(r)/r$  sees an $AdS_2$ at IR with the mass squared, $m^2=-\f{2B^2_2}{\rho^2+B^2_2(1+B^2_3)}$, which goes below the BF bound when the following  condition is met: $\f{\rho^2}{B^2_2}+B^2_3 < 7 $ (for a very specific to $B_3$, i.e. $B_3=0$, we get back the result of \cite{jkst}). From this it just follows trivially that, even for zero charge density case, we can have an instability, but to have a parameter that can control such an instability  requires it's better to have at least two nonvanishing magnetic fields, which is there in the above formula to mass. However, at UV the same scalar field sees an $AdS_4$ spacetime whose mass squared is above the BF bound  $m^2=-2\f{(1-B^2_3)}{1+B^2_3}$ for all real valued $B_3$.

For the other configuration, namely the D5-D5 brane system, the field $y(r)/r$ sees an $AdS_2$ spacetime at IR with the mass squared: 
$m^2=-\f{B^2_2}{\rho^2+B^2_2}$. Now, for the choice $\f{\rho}{B_2} <\sqrt{3}$, the scalar field not only becomes tachyonic as in the D3-D5 case; rather it goes below the BF bound, suggesting an instability. The same scalar field at UV rather sees an $AdS_3$ spacetime with mass squared $m^2=-1$, which just saturates the BF bound.

In another example, the D3-D7 system with zero charge density but with two magnetic fields, we see that the scalar field $y(r)/r$ enjoys an $AdS_3$ spacetime at IR with mass squared $m^2=-\f{3+B^2_3}{1+B^2_3}$, which violates the BF bound for any real value to $B_3$. Suggesting an instability,  this configuration somehow does not  display the necessary feature of a BKT-type transition. The plausible reason could be there is not any dimensionless parameter available like $\nu$  so as to  control such a violation of BF bound, rather the violation occurs generically.

Here we show the BKT type of transition by computing explicitly the 1-pt function as is done in \cite{jkst}, in particular, the condensate associated to the break down of the global rotational R symmetry: for the D3-D5 system it is the chiral symmetry $SU(2)_R$ and for D5-D5 it is a $U(1)_R$ symmetry. In both cases these symmetries are part of the R symmetry. In doing the calculation, one sets the source to zero, i.e., in the asymptotic expansion to the bulk field, which essentially means calculating the condensate for zero bare quark mass.
The result of this study can be summarized as follows:  (1)~There exists a BKT type of phase transition in 3+1 dimensional field theory, (2)~ as well as in 2+1 dimension but with zero charge density (apart from those found in \cite{jkst}) which means the BKT type of phase transition does not necessarily require an electric field, i.e., one can see these type of transitions only with the several magnetic fields.  (3)~ The potential energy in the linear regime has an inverse  power law type of behavior, and  (4) we see a BKT type of phase transition only when the spacetime admits an $AdS_2$ at IR with a dimensionless parameter $\nu$.

In the Dp-Dq system with electric  and several magnetic fields turned on, the asymptotic spacetime generically does not have the appropriate  behavior to be identified with the AdS spacetime. However, one can make appropriate field  redefinitions to bring it to the  AdS form all the time.

The organization of the paper is as follows. In section 2, we shall determine which are the Dp-Dq systems that are allowed, and in the next  section we shall  describe these systems by turning on  different kinds of magnetic fields. In the rest of the sections, we shall determine the condition for which there occurs a violation to the BF bound on a case by case basis and finally compute the 1-pt function for a few cases, and then conclude. In  Appendix A, we shall give a brief overview of seeing a BKT-type transition with $1/r^2$ potential. In Appendix B, we shall find the potentials seen in the D3-D5, D3-D7, and D5-D5 probe brane systems. In Appendix C, we shall calculate the contribution of the Chern-Simon part of the action  for these cases, also for D2-D4, D4-D6, and D5-D7 systems. In Appendix D, we shall give some properties of the probe brane solution in the D3-D5 system.

\section{Dp-Dq system}

The way we shall proceed to address the Dp-Dq brane system is by considering the Dq branes as probe branes in the background generated by the Dp branes in the quenched limit as prescribed in \cite{kk}.

In the large $N$  and large 'tHooft coupling limit, the coincident Dp branes are described as \cite{imsy}

\bea\label{dp_brane}
ds^2&=&f^{-\f{1}{2}}(r,y)[-dt^2+dx^2_1+\cdots+dx^2_p]+f^{\f{1}{2}}(r,y)[dr^2+r^2 d\Omega^2_n+dy^2+y^2 d\Omega^2_{7-p-n}],\nn
e^{\Phi}&=&g_s f^{\f{3-p}{4}}(r,y),~~~C_{p+1}=\f{1}{f(r,y)}dt\w dx_1\w \cdots\w dx_p,~~~f(r,y)=\f{R^{7-p}}{(r^2+y^2)^{\f{7-p}{2}}}
\eea 

The Dq brane is extended along the first $d$ spatial directions of  the Dp brane world volume  and wrapped over the $n$ dimensional sphere, $S^n$, and also extended along the  $r$ direction. So, counting the total number of spatial directions of the Dq brane is $q=d+1+n$. Now if we count the number of 
directions along which we have  the Neumann-Dirichlet 
boundary conditions,

\be
\# ND+\# DN\equiv {\bar \nu}=n+1+p-d
\ee

Now, we shall analyze  case by case,  i.e., for different choice to $p,~q,~n$ and $d$ in detail. In order to proceed, let us look for the situation where the $\# ND+\# DN=4$, i.e., $n+p=d+3$. So we shall restrict ourselves to supersymmetry preserving brane configuration in the absence of external fields. 

Starting with type IIA, for $p=0$, there exists only one   possibility that is $(q=4,~d=0,~n=3)$. 

For $p=2$, there occurs three different scenarios where this condition can be met. Those are $(q=2,~n=1,~d=0),~ (q=4,~n=2,~d=1),~(q=6,~n=3,~d=2)$, which means $q=2n$ and $d=n-1$ for $n=1,~2,~3$.

For $p=4$, there occurs four different scenarios where this condition can be met. Those are $(q=2,~n=0,~d=1),~ (q=4,~n=1,~d=2),~(q=6,~n=2,~d=3),~(q=8,~n=3,~d=4)$, which means $q=2d$ and $n=d-1$ for $d=1,~2,~3,~4$.

Now moving on to the type IIB case, there occurs similar kinds of scenarios, for example, for the $p=2$ case both $n$ and $d$
satisfy $n=d$ and $q=2n+1$ with $d=1,~2,~3$. For $p=5$,
$n=d-2$ and $q=2d-1$  with $d=2,~3,~4$. 

From this classification it just follows that in a single case one can turn on a maximum of three different kinds of magnetic fields for type IIA   that is for  D4-D8 for which $(n=3,~d=4)$. 

For   type IIB case, one can turn on a maximum of three different kinds of magnetic fields but only in one case. It happens for the D5-D7 system with $p=5$ and $q=7$ for which $(n=2,~d=4)$.

From this complete list of analyses, it follows that we can turn on a maximum of two different kinds of magnetic fields along the world volume of some of the Dp branes and only one kind of magnetic field along the  direction perpendicular to the Dp brane.

In the probe brane analysis, typically there occurs two different kinds of embedding functions; one parallel to the Dp brane world volume, which we shall denote as linear embedding (LE) and the second perpendicular to Dp brane, which we shall denote as the angular embedding (AE). In all of our discussion, we shall be  dealing with the latter type only, and keeping  the former type to future. 

In summary for type IIA, there exists seven possibilities:

\be
\begin{tabular}{||l|l|l|l|||l|l|l|l||}
\hline
Dp  & $(q,~n,~d)$ & embedding & $(B_1,B_2,B_3)$& Dp  & $(q,~n,~d)$ & embedding & $(B_1,B_2,B_3)$\\ \hline
D2 & $(2,~1,~0)$ & LE/AE & (0,0,0)&D4 & $(2,~0,~1)$ & LE/AE & $(0,0,0)$ \\ \hline
D2 & $(4,~2,~1)$ & LE/AE & $(0,0,B_3)$&D4 & $(4,~1,~2)$ & LE/AE & $(0,B_2,0)$ \\ \hline
D2 & $(6,~3,~2)$ & AE & $(0,B_2,B_3)$ &D4 & $(6,~2,~3)$ & LE/AE & $(0,B_2,B_3)$\\ \hline 
&&&&D4 & $(8,~3,~4)$ & AE & $(B_1,B_2,B_3)$ \\ \hline\hline\hline
\end{tabular}
\ee

and for type IIB, there exists six different possibilities:
\be
\begin{tabular}{||l|l|l|l|||l|l|l|l||}
\hline
Dp  & $(q,~n,~d)$ & embedding & $(B_1,B_2,B_3)$& Dp  & $(q,~n,~d)$ & embedding & $(B_1,B_2,B_3)$\\ \hline
D3 & $(3,~1,~1)$ & LE/AE & $(0,0,0)$&D5 & $(3,~0,~2)$ & LE/AE & $(0,B_2,0)$ \\ \hline
D3 & $(5,~2,~2)$ & LE/AE & $(0,B_2,B_3)$&D5 & $(5,~1,~3)$ & LE/AE & $(0,B_2,0)$ \\ \hline
D3 & $(7,~3,~3)$ & AE & $(0,B_2,B_3)$ &D5 & $(7,~2,~4)$ & LE/AE & $(B_1,B_2,B_3)$\\ \hline\hline\hline
\end{tabular}
\ee

For a given Dp-Dq system depending on the choice to $n$ and $d$, we can turn on zero magnetic field, one magnetic field, two magnetic fields, or three magnetic fields along with the electric field along the world volume of the probe Dq brane. As an example, for the D3-D5 system, we can turn on a maximum of two different magnetic fields, one along the common direction to the D3-D5 system and the other on the $S^2$ of the D7 brane world volume that it wraps.

\section{One magnetic field}

To set the tone, let us turn on a magnetic field of one kind  along the $x_{p-1}, x_p$ plane, which is parallel to the Dp brane and the electric field is along the $t,~r$ plane. 
Of course, one could have turned it along the $S^2$ of $S^n$ instead of the common world volume direction of the Dp-Dq system. We shall be dealing with the general case latter.
So, the 2-form field strength has the structure
\be\label{1_b_field}
F_2=-A'_0(r) dt\w dr+B dx_{d-1}\w dx_d.
\ee

Writing down the structure of the electric and magnetic fields as well as the brane directions,

\be
\begin{tabular}{l*{5}{c}r}
                & t & [$x_i: 1,~\cdots,p$] & r & $\Omega_n$ & y  &  $\Omega_{7-p-n}$\\
\hline
Dp              & x & [ $1,~\cdots,p$] &   &   &    &   \\

Dq              & x & [$ 1,~\cdots,d$] & x & x &   &    \\

E-field         & x &                     & x &   &   &     \\

B-field         &  & [$ d-1,d$]        &  &     &  &    \\
\hline
\end{tabular}
\ee
\\

The dynamics of  the $N_f$ Dq brane is governed by the DBI and CS action,
\be
S=-T_{Dq}N_f\int e^{-\Phi}\sqrt{-([g]_{ab}+F_{ab})}+\mu_{Dq}\int [C_{p+1}]\w e^F,
\ee 
where [~] denotes the pullback of the bulk fields onto the world volume of the Dq brane. The induced metric on the Dq brane is
\be
ds^2=f^{-\f{1}{2}}(r)[-dt^2+dx^2_1+\cdots+dx^2_d]+f^{\f{1}{2}}(r)[dr^2(1+y'(r)^2)+r^2 d\Omega^2_n]
\ee

The symmetry of the Dp brane is $SO(1,p)\times SO(9-p)_R$ but
the induced metric on the Dq brane enjoys lesser symmetry,   $SO(1,d)\times SO(p-d)\times SO(n+1)_R\times SO(8-p-n)_R$. Typically, the breaking of the global symmetry corresponds to the breaking  of $SO(8-p-n)_R$, where the subscript $R$ corresponds to the R symmetry.

On computing the determinant, the action becomes
\be
S=-{\cal N}\int dr r^n\sqrt{1-A'^2_0(r)+y'^2(r)}\sqrt{1+\f{B^2 R^{7-p}}{(r^2+y(r)^2)^{\f{7-p}{2}}}},
\ee  
where ${\cal N}=T_{Dq}N_f V_{d+1}\omega_n\sim \lambda^{\f{q-3}{4}}N_f N_c$. ~$V_{d+1}$ is the volume of the space $R^{1,d}$ and  $\omega_n$ is the volume of the sphere $S^n$.

Since the action does not depend on the zeroth component of the gauge field $A_0$, this implies the corresponding momentum is constant, which is nothing but the charge density associated to the U(1) gauge field.  By performing the Legendre transformation, we end up with an action without the derivative of the gauge field, 
\be
S_L=-{\cal N}\int dr \sqrt{1+y'^2(r)}\sqrt{\rho^2+\f{r^{2n}}{(r^2+y^2)^{\f{(7-p)(n+p-d-3)}{4}}}\Bigg(1+\f{B^2}{(r^2+y^2)^{\f{(7-p)}{2}}}\Bigg)},
\ee 
where $d={\cal N}\rho$.

If we want the Dp-Dq brane configuration to preserve some amount of supersymmetry before turning on the U(1) gauge field, then we need to set 
\be
n+p=d+3
\ee
under which the action reduces to 

\be
S_L=-{\cal N}\int dr \sqrt{1+y'^2(r)}\sqrt{\rho^2+r^{2n}+\f{B^2 r^{2n}}{(r^2+y^2)^{\f{(7-p)}{2}}}}.
\ee 

Let us also impose a further restriction on the choice to $p$ and $n$ such that both the charge density $\rho$ and magnetic field $B$ have the same dimension \cite{jkst},
\be
2n+p=7,
\ee
which is nothing but demanding that the dual  field theory lives on a maximum of four spatial directions. This then reduces the action to 
\be
S_L=-{\cal N}\int dr \sqrt{1+y'^2(r)}\sqrt{\rho^2+r^{2n}+\f{B^2 r^{2n}}{(r^2+y^2)^n}},
\ee 
whose quadratic fluctuation around the trivial solution $y=0$, results in 
\be\label{quadratic_action_1_b}
L\sim -\f{{\cal N}}{2}\sqrt{\rho^2+r^{2n}+B^2}~y'^2(r)+\f{n}{2}\f{{\cal N}B^2~y^2(r)}{r^2\sqrt{\rho^2+r^{2n}+B^2}}
\ee 

Analyzing this action results in the following possibilities of   Dp-Dq brane configurations where the Dq brane is wrapped on an $S^n$ and the field theory lives on a $d+1$ dimension:
 
\be
\begin{tabular} {l*{3}{c}r}
  n & d  & p & q    \\
\hline
  0 & 4 &  7 & 5        \\
  1 & 3 &  5 & 5        \\
  2 & 2 &  3 & 5        \\
\hline
  3 & 1 &  1 & 5        \\
  4 & 0 &  -1 & 5        \\
\hline
\end{tabular}
\ee

From this table it just follows that the last two configurations are not allowed which just  follows  from our construction, i.e., we needed at least 
two spatial directions along the world volume of the Dq brane which are common to the Dp brane, so that we can turn on a magnetic field. We shall come back to the situation latter where the magnetic field is turned on along the direction perpendicular to the Dp brane.

It's only the first three configurations that are legitimate. The configuration for which the D5 brane  is wrapped on an $S^2$ is already worked out in \cite{jkst}, so  we left with D7-D5 and D5-D5  brane configurations. The color D7 brane does not admit a decoupling limit and, hence, the D7-D5 configuration is not  that useful.

From the quadratic action, eq(\ref{quadratic_action_1_b}), it follows that near to the boundary the field $y/r$ becomes a massive scalar field in $AdS_{n+2}$ with mass squared $m^2=-n$, which never violates the Breitenlohner-Freedman bound \cite{bf}. However, at IR the field 
$y/r$ becomes a massive scalar field in $AdS_{2}$ with mass squared $m^2=-\f{nB^2}{\rho^2+B^2}$ which violates the Breitenlohner-Freedman bound for 
\be
\f{\rho}{B} < \sqrt{4n-1}.
\ee

There occurs a violation to the BF bound for D5-D5 brane configuration and this is the only other allowed brane configuration apart from D3-D5 that needed further investigation, whether it can show a BKT type of transition or not.

\subsection{Two magnetic fields}

Let us turn on the second magnetic field along the direction perpendicular to the Dp brane, along with the previous fields,  i.e., 
\be\label{2_b_field_A}
F_2=-A_0' dt\w dr+B_2 dx_{d-1}\w dx_d+B_3 sin~\theta d\theta\w d\phi
\ee
with the background geometry as
\be
ds^2=f^{-\f{1}{2}}[-dt^2+dx^2_1+\cdots+dx^2_p]+f^{\f{1}{2}}[dr^2+r^2 d\Omega^2_2+r^2s^2_\theta s^2_\phi d\Omega^2_{n-2}+dy^2+y^2 d\Omega^2_{7-p-n}],
\ee
which means we shall be restricting ourselves to $n\geq 2$, so that we can turn on a magnetic field $B_2$ on $S^2$ that sits inside $S^n$. For notational convenience we have written $sin~\theta=s_{\theta}$ and $sin~\phi=s_{\phi}$.

Writing down the full structure of the electric and magnetic fields as well as the brane directions

\be
\begin{tabular}{l*{6}{c}r}
                & t & [$x_i: 1,~\cdots,p$] & r & $\Omega_2$ &$\Omega_{n-2}$ & y  &  $\Omega_{7-p-n}$\\
\hline
Dp              & x & [ $1,~\cdots,p$] &   &  &  &    &   \\

Dq              & x & [$ 1,~\cdots,d$] & x & x & x & &    \\

E-field         & x &                     & x &  & &   &     \\

B-field         &  & [$ d-1,d$]        &  &   x &  &  &    \\
\hline
\end{tabular}
\ee
\\

The DBI action then becomes
\be
S=-\omega_n T_q N_f \int f^{\f{n+p-d-3}{4}}r^{n-2}
\sqrt{1+y'^2-A'^2_0}\sqrt{(1+B^2_2 f)(r^4+B^2_3 f^{-1})},
\ee
where $\omega_n$ is the volume of $S^n$ and $T_q$ is the tension of the Dq brane. Now doing the Legendre transformation we ended up with
\be\label{action_b1_b2}
S_L=-\omega_n T_q N_f\int 
\sqrt{1+y'^2}\sqrt{\rho^2+r^{2n-4}(1+B^2_2 f)(r^4+B^2_3 f^{-1})}.
\ee

The charge density $d=\f{\delta S}{\delta A'_0}=\omega_n T_q \rho$. The quadratically fluctuated  action around $y=0$  yields
\bea\label{action_b1_b2_lin}
\f{L}{\omega_n T_qN_f}&\sim& -\f{1}{2} \sqrt{\rho^2+r^{2n}(1+B^2_2 r^{p-7})+B^2_3 r^{2n-4}(B^2_2+r^{7-p})}y'^2-\nn &&\f{(7-p)}{4}
\f{[B^2_3 r^{2n-p+1}-B^2_2 r^{2n+p-9}]y^2}{\sqrt{\rho^2+r^{2n}(1+B^2_2 r^{p-7})+B^2_3 r^{2n-4}(B^2_2+r^{7-p})}}.
\eea

Let us take the  D3-D5 brane configuration, which means setting $n=2$. Doing the analysis as before, near to boundary the field $y/r$ becomes massive in  $AdS_4$ with $m^2=-2$, which lies within the Breitenlohner-Freedman bound but at IR, the field $y/r$ becomes massive and sees $AdS_2$ with 
\be
m^2=-\f{2B^2_2}{\rho^2+B^2_2+B^2_2B^2_3}.
\ee

So, the field $y/r$ can violate the Breitenlohner-Freedman bound at IR, if the following condition is met:
\be
\bigg(\f{\rho}{B_2}\bigg)^2+B^2_3 < 7.
\ee

Now going by the logic that violation of the BF bound can generate an instability suggests that there can even be a quantum phase transition for zero charge density, but for that to occur we needed  at least two nonzero magnetic fields  in which case the strength of the second magnetic field must obey $B_3 < \sqrt{7}$. The requirement of the second magnetic field will be clear when we compute the condensate, so that we can have a parameter on which the condensate depends, e.g., $B_3$ for zero charge density case. 

\subsection{Three magnetic fields}

Let us turn on yet another  magnetic field along the direction parallel to the Dp brane, along with the previous configurations as suggested in section 2, 
\be
F_2=-A_0' dt\w dr+B_1 dx_{d-3}\w dx_{d-2}+B_2 dx_{d-1}\w dx_{d}+B_3 sin~\theta d\theta\w d\phi
\ee
 
In this case the brane configuration is summarized  as follows:

\be
\begin{tabular}{l*{6}{c}r}
                & t & [$x_i: 1,~\cdots,p$] & r & $\Omega_2$ &$\Omega_{n-2}$ & y  &  $\Omega_{7-p-n}$\\
\hline
Dp              & x & [ $1,~\cdots,p$] &   &  &  &    &   \\

Dq              & x & [$ 1,~\cdots,d$] & x & x & x & &    \\

E-field         & x &                     & x &  & &   &     \\
$B_1$-field         &  & [$ d-3,d-2$]        &  &    &  &  &    \\
$B_2$-field         &  & [$ d-1,d$]        &  &    &  &  &    \\
$B_3$-field         &  &         &  &   x &  &  &    \\
\hline
\end{tabular}
\ee
\\

On evaluation of the DBI action,
\be
S=-\omega_n T_q N_f\int f^{\f{n+p-d-3}{4}}r^{n-2}
\sqrt{1+y'^2-A'^2_0}\sqrt{(1+f B^2_1 )(f^{-1}+B^2_2 )(fr^4+B^2_3 )},
\ee
where $\omega_n$ is the volume of $S^n$ and $T_q$ is the tension of the Dq brane. Again, let us demand that the configuration preserves some amount of supersymmetry before turning on the U(1) gauge field, suggests to set $n+p=d+3$. As previously, doing the Legendre transformation we end up with
\be\label{action_b1_b2_b3}
S_L=-\omega_n T_qN_f \int 
\sqrt{1+y'^2}\sqrt{\rho^2+r^{2n-4}(1+f B^2_1 )(f^{-1}+B^2_2 )(fr^4+B^2_3 )}.
\ee
The charge density $d=\f{\delta S}{\delta A'_0}=\omega_n T_q \rho$. The quadratically fluctuated  action around the trivial solution $y=0$, with the choice $R=1$,  yields

\bea\label{quadratic_action_b1_b2_b3}
&&\f{L}{\omega_n T_qN_f}\sim -\f{1}{2}\sqrt{\rho^2+r^{2n-4}(1+B^2_1 r^{p-7})(r^{7-p}+B^2_2)(r^{p-3}+B^2_3)}y'^2-\nn &&
\f{(7-p)}{4}\f{[B^2_3r^{2n-p+1}-B^2_1B^2_2B^2_3r^{2n+p-13}-(B^2_1+B^2_2)r^{2n+p-9}-
2B^2_1B^2_2r^{2n+2p-16}]}{\sqrt{\rho^2+r^{2n-4}(1+B^2_1 r^{p-7})(r^{7-p}+B^2_2)(r^{p-3}+B^2_3)}}y^2\nn
\eea

Having obtained the most general fluctuated action, now we can proceed to check on a case by case basis where there could be a possibility to violate  the Breitenlohner-Freedman bound. There naturally arises two different situations to do the analysis: those are with zero density and nonzero density along with the magnetic fields.

\section{Zero density }

For zero density, eq(\ref{action_b1_b2_b3}) reduces to
\be\label{action_b1_b2}
S_L=-\omega_n T_q N_f\int r^{n-2}
\sqrt{1+y'^2}\sqrt{\bigg(1+\f{B^2_1}{(r^2+y^2)^{\f{7-p}{2}}}\bigg)\bigg((r^2+y^2)^{\f{7-p}{2}}+B^2_2 \bigg)\Bigg(\f{r^4}{(r^2+y^2)^{\f{7-p}{2}}}+B^2_3 \Bigg)},
\ee
where we have used $f=(r^2+y^2)^{(p-7)/2}$. Let us  concentrate on the simplest situation for which all the magnetic fields  vanish, i.e., we are  going to analyze the D4-D2  brane configuration and this give an 
$AdS_{2}$ space,  for the massless scalar field $y/r$ at IR. 
For D2-D2 and D3-D3 this gives an $AdS_3$ space for the field $y/r$ at IR with mass $m^2=-1$, which saturates the 
Breitenlohner-Freedman bound.

With two vanishing magnetic fields like $B_1=0=B_3$ and one nonvanishing field $B_2\neq 0$, this gives a brane configuration like  D4-D4, D5-D3, and D5-D5. The D4-D4 gives an $AdS_{2}$ at IR for the field $y/r^{\f{5}{4}}$ with mass as $m^2=-\f{3}{16}$ and $m^2=-\f{19}{16}$, respectively. But for the D5-D3 system it gives $AdS_3$
for the field $y/r^2$ with $m^2=-1$. For  D5-D5, it gives $AdS_2$ with $m^2=-1$ which saturates the BF bound. 

If the vanishing magnetic fields are $B_1=0=B_2$, then it corresponds to only one configuration that is D2-D4. This gives an $AdS_2$ for massless field $y/r$  at IR for both zero and nonzero density.

Let us consider the situation where at least two magnetic fields are nonzero, in particular, for $B_1=0$. The brane configurations that comes under this are D2-D6, D4-D6, D3-D5, and D3-D7.
For the D2-D6  it becomes $AdS_{2}$ with $m^2=-\f{43}{16}$ for the field $y/r^\f{3}{4}$ at IR. For D4-D6   it gives $AdS_{2}$ with massless scalar  field for $y/r$ at IR,  whereas for D3-D5 it becomes $AdS_2$ with $m^2=\f{-2}{1+B^2_3}$ for the field $y/r$ at IR which can violate the BF bound when  the condition $B_3 < \sqrt{7}$ is met. For D3-D7, it becomes $AdS_3$ for the field $y/r$ with $m^2=-\f{3+B^2_3}{1+B^2_3}$ at IR which also violates the BF bound for any value to $B_3$.

When none of the magnetic field vanishes, there arises two different situations D5-D7 and D4-D8. For the D5-D7 case, it gives $AdS_3$ with $m^2=-1$ for the field $y/r^2$ at IR for both zero and nonzero density.

Summarizing the behavior at IR for the zero density case,
\be
\begin{tabular}{||l|l|l|l|l|l||}
\hline
$(Dp,~Dq,~n,~d)$ & $(B_1,~B_2,~B_3)$ & $m^2$ & $AdS_m$ & field & BF bound  \\ \hline
$(D4,~D2,~0,~1)$ & $(0,0,0)$ & $0$ & $AdS_2$ & $y/r$ & Preserves  \\ \hline
$(D2,~D2,~1,~0)$ & $(0,~0,~0)$ & $-1$ & $AdS_3$ & $y/r$ & Saturates  \\ \hline
$(D3,~D3,~1,~1)$ & $(0,~0,~0)$ & $-1$ & $AdS_3$ & $y/r$ & Saturates  \\ \hline
\end{tabular}
\ee

Note that with zero  electric and magnetic fields the Dp-Dq brane configuration is supersymmetric and, hence, should be stable all the time, which is consistent with the fact that the field $y$ or $y/r^n $ for any $n$ never violates the BF bound:
\be\label{table_zero_density}
\begin{tabular}{||l|l|l|l|l|l|l||}
\hline
$(Dp,~Dq,~n,~d)$ & $(B_1,B_2,B_3)$ & $m^2$ & $AdS_m$ & field & BF bound& Condition ?  \\ \hline
$(D4,~D4,~1,~2)$ & $(0,~B_2,~0)$ & $-\f{19}{16}$ & $AdS_{2}$ & $y/r^{\f{5}{4}}$ & Violates& Generically  \\ \hline
$(D5,~D3,~0,~2)$ & $(0,~B_2,~0)$ & $-1$ & $AdS_3$ & $y/r^2$ & Saturates & \\ \hline
$(D5,~D5,~1,~3)$ & $(0,~B_2,~0)$ & $-1$ & $AdS_2$ & $y/r$ & Preserves & \\ \hline\hline\hline
$(D2,~D4,~2,~1)$ & $(0,0,B_3)$ & $0$ & $AdS_{2}$ & $y/r$ & Preserves&   \\ \hline\hline\hline
$(D2,~D6,~3,~2)$ & $(0,B_2,B_3)$ & $-\f{43}{16}$ & $AdS_{2}$ & $y/r^{\f{3}{4}}$ & Violates& Generically  \\ \hline
$(D4,~D6,~2,~3)$ & $(0,B_2,B_3)$ & $0$ & $AdS_{2}$ & $y/r$ & Preserves&   \\ \hline
$(D3,~D5,~2,~2)$ & $(0,B_2,B_3)$ & $-\f{2}{1+B^2_3}$ & $AdS_2$ & $y/r$ & Violates &$B_3<\sqrt{7}$ \\ \hline
$(D3,~D7,~3,~3)$ & $(0,B_2,B_3)$ & $-\f{3+B^2_3}{1+B^2_3}$ & $AdS_3$ & $y/r$ & Violates &Generically \\ \hline\hline\hline
$(D4,~D8,~3,~4)$ & $(B_1,B_2,B_3)$ & $-\f{19}{16}$ & $AdS_{2}$ & $y/r^{\f{5}{4}}$ & Violates &Generically   \\ \hline
$(D5,~D7,~2,~4)$ & $(B_1,B_2,B_3)$ & $-1$ & $AdS_3$ & $y/r^2$ & Saturates & \\ \hline\hline\hline
\end{tabular}
\ee

This calculation is done only by looking at the DBI action, but as we know the Chern-Simon term is also important and its   contribution  can change some of the results.
In some cases we calculate the contribution of the Chern-Simon action and this is presented in Appendix C.

Here we shall not be dealing with configurations like D4-D4, D2-D6, and D4-D8, even though they violates the BF  bound at IR generically.   So, we are left with the   configurations of D3-D5 and D3-D7 whose 1-pt function we shall explore.

\section{Calculation of 1-pt function}

In this section we shall present the calculation of  the 1-pt function for  systems that violates the BF bound at IR, mostly following \cite{dss}, \cite{kos} and \cite{jkst}.
It is interesting to note that the violation to the BF bound occurs only at IR, not at UV, which suggests that we can do the calculation of the 1-pt function following the algorithm of \cite{dss} and \cite{kos}.

Recalling from \cite{kw}, it is suggested that, when the mass of the scalar field stays  above the BF bound, the $\Delta_+$ branch is the legitimate branch all the time, apart from the $\Delta_-$ branch which comes into the picture only when the mass of the scalar field stays within a specific range. Close to the boundary $r\rightarrow \infty$, the scalar field takes the structure 
\be\label{boundary_phi}
\Phi(r,x_i)\rightarrow r^{\Delta-d}\phi_0(x_i)+r^{-\Delta}A(x_i),
\ee

where $\phi_0(x_i)$ is interpreted as the source and $A(x_i)$ as the vev of the operator dual to the scalar field $\phi$ or to the operator that $\phi_0$ couples to on the boundary. But this interpretation of source and vev gets interchanged when we go over to that specific mass range for which  both the $\Delta_+$ and $\Delta_-$ branches are allowed. In what follows, we shall restrict ourselves to the $\Delta_+$ branch and interpret $A(x_i)$ as the vev, i.e., the condensate.

We know that the 1-pt function for the bulk field $\Phi$ is nothing but the momentum associated to the field, evaluated at the boundary, which means
\be
<O>=\f{\delta S}{\delta \Phi(r)}=\f{\p L}{\p (\p_r\Phi)}=\pi|_{boundary}.
\ee 

To compute the 1-pt function, as an example, let us take the 
D3-D5 system with $B_1=0=B_3$;  the fluctuated action around $y=0$  follows from eq(\ref{quadratic_action_b1_b2_b3}). Computing the momentum, 
\be
\pi=-\omega_2 T_5 N_f\sqrt{\rho^2+B^2+r^4}y'.
\ee 

Now, using the asymptotic solution to $y$, 
\be
y=y_0+y_1/r+\cdots,
\ee
where $y_0$ and $y_1$ are some constants, gives 
 the momentum at the boundary, which  at $r=\infty$, results
\be
\pi=\omega_2 T_5 N_fy_1=<O>.
\ee

So, we see that it is the asymptotic expansion of the bulk field that gives us the desired result to the  calculation of  the 1-pt function. In the next section, we shall be dealing with  the D3-D5 case  in more detail.

\subsection{D3-D7 system}

We see from the table, eq(\ref{table_zero_density}), that at zero density the D3-D7 system violates the BF bound generically but only for the nonzero choice to $B_2$ and this violation is true irrespective of any value to $B_3$. So, let us set $B_3=0$ for simplicity. The action is
\be
S=-\omega_3 T_7N_f\int r \sqrt{1+y'^2}\sqrt{r^4+\f{B^2_2r^4}{(r^2+y^2)^2}}
\ee

The linearized equation of motion that follows around the trivial solution $y=0$ is
\be
r^2(B^2_2+r^4)y''+r(B^2_2+3r^4)y'+2B^2_2y=0
\ee
The solution with two arbitrary constants, $c_1$ and $c_2$ are
\be
y=c_1 ~cos\bigg(\f{1}{\sqrt{2}}~Log\bigg(\f{B^2_2+B_2\sqrt{B^2_2+r^4}}{r^2}\bigg)\bigg)+c_2 ~sin\bigg(\f{1}{\sqrt{2}}~Log\bigg(\f{B^2_2+B_2\sqrt{B^2_2+r^4}}{r^2}\bigg)\bigg)
\ee

It looks as if we can have the BKT phase transition in this case too, but that is not quite correct. The correct way is to look at the behavior of the condensate and from it one can say whether it will  show the BKT transition or not.  

The asymptotic behavior of \be
Log\bigg(\f{B^2_2+B_2\sqrt{B^2_2+r^4}}{r^2}\bigg)\bigg)\rightarrow Log~B_2+\f{B_2}{r^2}+\cdots,
\ee
so the solution to $y$ becomes
\be
y=[c_1~cos(Log~B_2)+c_2~sin(Log~B_2)]+\f{B_2}{r^2}\bigg[-c_1~sin(Log~B_2)+c_2~cos(Log~B_2)\bigg]+{\cal }O(1/r^3),
\ee
which is what follows because at UV it becomes $AdS_3$ for the massless scalar field $y$. Choosing for the masslessness  of quarks means setting  the order $r^0$ term in $y$ to zero, which results in $y=\sigma/r^2$, where the condensate 
\be
\sigma=B_2 \bigg(\f{c^2_1+c^2_2}{c_2}\bigg)~cos(Log~B_2).
\ee

The structure of the condensate is not of the $e^{-\f{c}{\sqrt{\nu_c-\nu}}}$ type, which means there will not be any BKT transition for the D3-D7 system, It  is suggested in 
\cite{jkt} and \cite{fjrv} that this transition is second order. The plausible reason of not seeing a BKT-type transition  is  that  the scalar field $y/r$ sees an $AdS_3$ spacetime  instead of $AdS_2$ at IR. Just to recall from the study of \cite{bkt},  the  BKT-type transition occurs only for two dimensional (Euclidean) spaces.  

Since the condensate is oscillatory means for  specific choice to $B_2$, we can have zero condensate. For $B_2=0$, the condensate vanishes suggesting at zero temperature with zero density and zero quark mass there should not be any chiral symmetry breaking.

\section{Non-zero density}
Let us look at the nonzero density case and see if some of the results can get modified   in the presence of charge density or not:
\be
\begin{tabular}{||l|l|l|l|l|l||}
\hline
$(Dp,~Dq,~n,~d)$ & $(B_1,~B_2,~B_3)$ & $m^2$ & $AdS_m$ & field & BF bound  \\ \hline
$(D4,~D2,~0,~1)$ & $(0,0,0)$ & $0$ & $AdS_2$ & $y/r$ & Preserves  \\ \hline
$(D2,~D2,~1,~0)$ & $(0,~0,~0)$ & $0$ & $AdS_2$ & $y/r$ & Preserves  \\ \hline
$(D3,~D3,~1,~1)$ & $(0,~0,~0)$ & $0$ & $AdS_2$ & $y/r$ & Preserves  \\ \hline
\end{tabular}
\ee

For a nonzero electric field but  zero magnetic field, we saw that the  brane configuration does preserve the BF bound. It should not be thought of as if the system possesses some amount of supersymmetry, which should in fact be checked by doing the Kappa symmetry preserving calculation. But we are not interested in that at the present time. Different intersecting brane configurations can be summarized as follows:

\be\label{table_density}
\begin{tabular}{||l|l|l|l|l|l|l||}
\hline
$(Dp,~Dq,~n,~d)$ & $(B_1,B_2,B_3)$ & $m^2$ & $AdS_m$ & field & BF bound& Condition ?  \\ \hline
$(D4,~D4,~1,~2)$ & $(0,~B_2,~0)$ & $-\f{19}{16}$ & $AdS_{2}$ & $y/r^{\f{5}{4}}$ & Violates& Generically \\ \hline
$(D5,~D3,~0,~2)$ & $(0,~B_2,~0)$ & $-1$ & $AdS_3$ & $y/r^2$ & Saturates & \\ \hline
$(D5,~D5,~1,~3)$ & $(0,~B_2,~0)$ & $-\f{B^2_2}{\rho^2+B^2_2}$ & $AdS_2$ & $y/r$ & Violates &$\f{\rho}{B_2}<\sqrt{3}$ \\ \hline\hline\hline
$(D2,~D4,~2,~1)$ & $(0,0,B_3)$ & $0$ & $AdS_{2}$ & $y/r$ & Preserves&   \\ \hline\hline\hline
$(D2,~D6,~3,~2)$ & $(0,B_2,B_3)$ & $0$ & $AdS_{2}$ & $y/r$ & Preserves& \\ \hline
$(D4,~D6,~2,~3)$ & $(0,B_2,B_3)$ & $0$ & $AdS_{2}$ & $y/r$ & Preserves&   \\ \hline
$(D3,~D5,~2,~2)$ & $(0,B_2,B_3)$ & $\f{-2B^2_2}{\rho^2+B^2_2(1+B^2_3)}$ & $AdS_2$ & $y/r$ & Violates &$\f{\rho^2}{B^2_2}+B^2_3<7$ \\ \hline
$(D3,~D7,~3,~3)$ & $(0,B_2,B_3)$ & $0$ & $AdS_2$ & $y/r$ & preserves & \\ \hline\hline\hline
$(D4,~D8,~3,~4)$ & $(B_1,B_2,B_3)$ & $-\f{19}{16}$ & $AdS_{2}$ & $y/r^{\f{5}{4}}$ & Violates& Generically \\ \hline
$(D5,~D7,~2,~4)$ & $(B_1,B_2,B_3)$ & $-1$ & $AdS_3$ & $y/r^2$ & Saturates & \\ \hline\hline\hline
\end{tabular}
\ee

On comparing the zero density case with the nonzero density case, we see that the D5-D5 brane configuration which was preserving the BF bound in the former case is now violating it. Similarly, the D3-D7 system which was violating the BF bound at zero density case is now preserving the BF bound. Hence, the configurations that we shall be dealing with are D3-D5 and D5-D5.

\section{Broken phase: $D3-D5-E-B_2-B_3$}

In this section we shall calculate the condensate or the 1-pt function associated to the operator dual to the scalar field $y$. Generically, it is very difficult to solve the nonlinear equations of motion that follows from eq(\ref{action_b1_b2_b3}). So the approach will be the same as studied in \cite{jkst}. Numerically, the solution will be found in the $r=0$ region called the core region and it exhibits the same behavior as one gets solving the linearized equation of motion that follows from eq(\ref{quadratic_action_b1_b2_b3}) in the small $r$ region. This then will be compared to the solution that follows from solving the linearized equation of motion that follows from eq(\ref{quadratic_action_b1_b2_b3}) but in the large $r$ region. From the resulting solution we shall read out the condensate.  

In the core region, when the scalar field saturates the BF bound, the action becomes
\be\label{core_action}
S_{core}=-\omega_2 T_5 B_2N_f\int \sqrt{1+y'^2}\sqrt{7+\f{r^4}{(r^2+y^2)^2}}, 
\ee
whose numerical solution with the boundary condition $y(0)=0$ can be fitted to 
\be\label{sol_r_r0}
y=\sqrt{r}[-a_0+a_1 Log~r]=a_1\sqrt{r}Log\bigg(\f{r}{r_0}\bigg)
\ee 
for $r>>r_0$, here $a_0$ and $a_1$ are some constants with  finite valued  real numbers. 

The easier way to see the solution of the $log$ form is by looking at the action, eq(\ref{quadratic_action_b1_b2_b3}),  which gives the equation of motion of the form, when the mass of the  scalar field just saturates the BF bound,
\be\label{eom_r_r0}
y''+\f{y}{4r^2}=0.
\ee 

This equation as well as the action at the core, eq(\ref{core_action}), shows the presence of a scaling symmetry under which both $r$ and $y$ scales in the same way, which  left the form of the equation of motion
unchanged. So a general form of the solution at the core can be 
\be\label{sol_y_d3_d5_small_r}
y_\xi=a_1\sqrt{r\xi}Log\bigg(\f{r}{r_0}\bigg).
\ee

The solution to eq(\ref{eom_r_r0}) has the structure of eq(\ref{sol_r_r0}). Assuming the following choice of the boundary condition, $y(r_0)=y_0$ and $y'(r_0)=y_1$ in eq(\ref{sol_r_r0}) gives the result for $y_0=0$ that $a_0/a_1=Log~r_0$ and the velocity $y_1=\sqrt{\f{a_0a_1}{r_0~Log~r_0}}$. 

In order to find  the far away solution, we shall solve the equation of motion that follows from eq(\ref{quadratic_action_b1_b2_b3}) and it gives the solution as
\be
y\sim c_+~f_++c_-~f_-,
\ee
where $c_{\pm}$ are two arbitrary constants and $f_{\pm}$ are the two solutions of the second order differential equation, 
\be
f_{\pm}=u^{\f{1\pm i\alpha}{2}}{}_2F_1[\f{2-\beta\pm i\alpha}{8},\f{2+\beta\pm i\alpha}{8},1\pm\f{ i\alpha}{4},-u^4],
\ee
where 
\be\label{parameters_d3_d5}
\alpha=\sqrt{\f{\rho^2_c-\rho^2}{\rho^2+B^2_2(1+B^2_3)}},~~~\beta=\sqrt{\f{1+9B^2_3}{1+B^2_3}},~~~u=\delta ~r,~~~\delta=\bigg(\f{1+B^2_3}{\rho^2+B^2_2(1+B^2_3)}\bigg)^{1/4}
\ee
 ${}_2F_1[a,b,c,z]$ is the hypergeometric function and $\rho_c=B_2\sqrt{7-B^2_3}$.
The way to put the boundary condition is  to set the bare quark mass to zero and asymptotically 
\be
f_n=c_+~f_++c_-~f_-\longrightarrow \f{\sigma}{u^{(1+\beta)/2}}, ~~~u\rightarrow\infty,
\ee
where $y=-f_n/\delta^{\f{1+\beta}{2}}$. Denoting the large $u$  expansion of $f_{\pm}$,
\be
f_+\sim c_0 u^{-(1+\beta)/2}+c_1 u^{-(1-\beta)/2},~~~
f_-\sim {\tilde c}_0 u^{-(1+\beta)/2}+{\tilde c}_1 u^{-(1-\beta)/2}
\ee
and solving for the zero bare quark mass and the asymptotic behavior of $f_n$  fixes the two unknown coefficients as
\be
c_+=\f{\sigma {\tilde c}_1}{c_0{\tilde c}_1-c_1{\tilde c}_0},~~~c_-=\f{\sigma  c_1}{c_1{\tilde c}_0-c_0{\tilde c}_1}
\ee

The small $u$  expansion of $f_n$ becomes
\be
f_n\sim c_+ u^{(1+i\alpha)/2}+c_-u^{(1-i\alpha)/2}=\f{\sigma\sqrt{u}}{c_1{\tilde c}_0-c_0{\tilde c}_1}[{\tilde c}_1u^{i\alpha/2}-c_1 u^{-i\alpha/2}]
\ee
Now, using the small $\alpha$ limit allows us to rewrite
\be
f_n=-\sigma \f{Y\sqrt{u}}{\alpha}~sin~\bigg(\f{\alpha}{2}Log(u/u_1)\bigg),
\ee 
where $u_1=e^{2\f{X}{Y}}=e^{\f{\gamma+\Psi(\f{2+\beta}{8})}{4}}$  and 
\be
X=\f{{\tilde c}_1-c_1}{c_1{\tilde c}_0-c_0{\tilde c}_1}=\f{\Gamma[\f{2-\beta}{8}]\Gamma[\f{6-\beta}{8}]\bigg(\gamma+\Psi(\f{2+\beta}{8})\bigg)}{\Gamma[-\f{\beta}{4}]\bigg(\Psi(\f{2-\beta}{8})-\Psi(\f{2+\beta}{8})\bigg)}+{\cal O}(\alpha^2),
\ee
$i\f{Y}{\alpha}=\f{{\tilde c}_1+c_1}{c_1{\tilde c}_0-c_0{\tilde c}_1}$,
\be
Y=\f{8~\Gamma[\f{2-\beta}{8}]\Gamma[\f{6-\beta}{8}]}{\Gamma[-\f{\beta}{4}]\bigg(\Psi(\f{2-\beta}{8})-\Psi(\f{2+\beta}{8})\bigg)}+{\cal O}(\alpha^1),
\ee

where $\gamma$ is the Euler's constant and $\Psi(z)=\f{\Gamma'[z]}{\Gamma[z]}$, is the digamma function. This gives 
\be\label{sol_y_d3_d5_large_r}
y=-\delta^{(1+\beta)/2}f_n=\delta^{(2+\beta)/2}
\f{\sigma\sqrt{r}}{\alpha}Y ~sin~\bigg(\f{\alpha}{2}Log(r/{\tilde r}_1)\bigg),
\ee
where ${\tilde r}_1=\f{1}{\delta}e^{2X/Y}$. In order to match the core and the linear regime solutions, we demand the argument of the $sin$ function should be a multiple of $\pi$, which means evaluating at $r=\xi$ results in
\be
\xi \sim e^{-2\pi/\alpha}
\ee
and comparing eq(\ref{sol_y_d3_d5_small_r}) with eq(\ref{sol_y_d3_d5_large_r}) implies $\sigma \sim \sqrt{\xi}$, which suggests
\be\label{condensate_d3_d5_non_zero_rho}
\sigma \sim e^{-\pi/\alpha}=e^{-\f{{\tilde c}}{\sqrt{\rho_c-\rho}}}=e^{-\f{c}{\sqrt{\nu_c-\nu}}},
\ee
where 
\be\label{condensate_d3_d5_non_zero_rho1}
{\tilde c}=\pi\sqrt{\f{\rho^2+B^2_2(1+B^2_3)}{\rho_c+\rho}},~~~\nu=\f{\rho}{B_2\sqrt{1+B^2_3}},~~~c=\pi\sqrt{\f{\nu^2+1}{\nu+\nu_c}},~~~\nu_c=\f{\rho_c}{B_2\sqrt{1+B^2_3}}
\ee

From this formula, even though it looks as if turning on a magnetic field of $B_3$ kind simply rescales  either $B_2$ or $\rho$,  that is not quite correct. Let us recall that the mass of scalar field $y/r$ goes as $m^2=-\f{2B^2_2}{\rho^2+B^2_2(1+B^2_3)}$ at IR, which under rescaling of $B_2$ and $\rho$ do not reproduces the $B_3\rightarrow 0$ limit i.e.    $m^2=-\f{2B^2_2}{\rho^2+B^2_2}$. In fact one can write down the mass of the field $y/r$ at IR as
\be
m^2=
-\f{2B^2_2}{\rho^2+B^2_2(1+B^2_3)}=
-\f{2}{(1+B^2_3)(1+\nu^2)},
\ee
which contains two parameters $B_3$ and $\nu$, whereas the condensate in eq(\ref{condensate_d3_d5_non_zero_rho}) depends only on one parameter, $\nu$.

Let us rewrite eq(\ref{condensate_d3_d5_non_zero_rho}) for simplicity in the $B_3 \rightarrow 0$ limit  but in a different way; i.e., in terms of the conformal dimension $\Delta$ at IR. The mass squared $m^2=-\f{2B^2_2}{\rho^2+B^2_2}=-\f{2}{1+\nu^2}$ and using $\Delta(\Delta-1)=m^2$, because the field $y/r$ sees an $AdS_2$ at IR, we get 
\be
\sigma~\sim~ e^{-\f{C}{\sqrt{\Delta_c-\Delta}}},
\ee
where 
\be
C=\pi \sqrt{\f{\Delta_c(1-\Delta_c)}{\Delta+\Delta_c-1}},
\ee
which for $\Delta\sim\Delta_c$ becomes
\be
C=\pi\sqrt{\f{\Delta_c(1-\Delta_c)}{2\Delta_c-1}}
\ee

where $\Delta_c$ is determined from the equation $\Delta_c(\Delta_c-1)=m^2_c=-\f{2}{1+\nu^2_c}$.

\subsection{Broken phase: $D3-D5-B_2-B_3$ with $\rho=0$}

It looks from the analysis in \cite{jkst}   as if the presence of electric field is essential in order to see the BKT phase transition. However, this is not completely true,  which can be seen from this example. In this case we turned off the electric field but left untouched both the magnetic fields $B_2$ and $B_3$.

The action of the scalar field at the core region is precisely the same as is written for the $\rho\neq 0$ case.
In this case note that the BF bound is violated when the strength of the magnetic field $B_3$ goes below  $\sqrt  {7}$, i.e., obeys the condition $B_3 < \sqrt  {7} $ . Again, the solution at the core can be found numerically by putting the Dirichlet boundary condition at $r=0$ and the fitted function looks the same as for the $\rho\neq 0$ case.

This solution is going to be matched somewhere in the middle of the far away solution. The far away solution, again, has the same structure as for the $\rho\neq 0$ case, but with one difference, that is, 
\be\label{alpha_0}
\alpha =\sqrt{\f{(B^2_{3})_c-B^2_3}{1+B^2_3}},
\ee 
where $(B^2_{3})_c$ is the square of the strength of the critical magnetic field perpendicular the D3 brane below  which there is a violation of  the BF bound and whose value is $(B_3)_c=7$.

Proceeding as in the previous case for $\rho\neq 0$, we find the condensate goes as
\be\label{condensate_d3_d5_zero_rho}
\sigma ~\sim ~ e^{-\pi/\alpha}=e^{-\f{{\tilde c}}{\sqrt{(B_3)_c-B_3}}}=e^{-\f{c}{\sqrt{\nu_c-\nu}}},
\ee
where 
\be
{\tilde c}=\pi\sqrt{\f{1+B^2_3}{(B_3)_c-B_3}},~~~\nu=B_3,~~~c=\pi\sqrt{\f{\nu^2+1}{\nu_c+\nu}}.
\ee

One can see that the result of eq(\ref{condensate_d3_d5_zero_rho}) follows directly in the $\rho$ going to zero limit of eq(\ref{condensate_d3_d5_non_zero_rho}). However, $\nu$ simply does not follow directly from eq(\ref{condensate_d3_d5_non_zero_rho1}), but can be seen from the expression to $\alpha$ as written in eq(\ref{parameters_d3_d5}). This gives another clue that $B_3$ cannot be removed by a scaling to either $B_2$ or $\rho$. 

\subsection{D3-D5 system with Full action}

In this section we shall include the contribution of the Chern-Simon part of the action into the probe brane action. Following the result from Appendix C, we find the full action density to probe  the D5 brane with the notation  
 $\Sigma=\mu_54\pi N_f B_3$ as
\be
S=-\alpha\int \sqrt{x'^2_3f^{-1}+1+y'^2-A'^2_0}\sqrt{(f^{-1}+B^2_2)(fr^4+B^2_3)}+\Sigma \int \f{x'_3(r)}{f},
\ee
where $\alpha=T_5 4\pi N_f$,  the integration is over $r$, and the volume is that of $R^{1,2}$. Since the action does not depend on $A_0$ and $x_3$ means the corresponding momenta are constants,
\bea
d&\equiv& \f{\delta S}{\delta A'_0}=\alpha A'_0\f{\sqrt{(f^{-1}+B^2_2)(fr^4+B^2_3)}}{\sqrt{x'^2_3f^{-1}+1+y'^2-A'^2_0}},\nn
C&\equiv& \f{\delta S}{\delta x'_3}=\f{\Sigma}{f}-\alpha x'_3 f^{-1}\f{\sqrt{(f^{-1}+B^2_2)(fr^4+B^2_3)}}{\sqrt{x'^2_3f^{-1}+1+y'^2-A'^2_0}}
\eea

From this conserved quantity, it just follows that 
\be
A'^2_0 (Cf-\Sigma)^2=d^2 x'^2_3.
\ee 

Using this relation between $A'_0$ and $x'_3$, we arrive at the result 
\bea\label{sol-x3-A0}
A'^2_0&=&\f{d^2(1+y'^2)}{\alpha^2(f^{-1}+B^2_2)(fr^4+B^2_3)+\rho^2\alpha^2-f^{-1}(Cf-\Sigma)^2},\\
x'^2_3&=&\f{(Cf-\Sigma)^2(1+y'^2)}{\alpha^2(f^{-1}+B^2_2)(fr^4+B^2_3)+\rho^2\alpha^2-f^{-1}(Cf-\Sigma)^2},
\eea
where we have rewritten the charge density as $d=\alpha\rho$. Doing the Legendre transformation of  the action, results in
\bea
S_L&=&S-\int\f{\delta S}{\delta A'_0}A'_0-\int\f{\delta S}{\delta x'_3}x'_3\nn &=&-\int \sqrt{1+y'^2}\sqrt{\alpha^2(f^{-1}+B^2_2)(fr^4+B^2_3)+\rho^2\alpha^2-f^{-1}(Cf-\Sigma)^2}
\eea

Let us work in a specific choice to the momentum associated to $x_3$ that is $C=0$, in which case the Legendre transformed action becomes 
\be\label{lin_dbi_cs}
S_L=-\alpha \int \sqrt{1+y'^2}\sqrt{(f^{-1}+B^2_2)(fr^4+B^2_3)+\rho^2-f^{-1}\Sigma^2/\alpha^2}
\ee

This choice to momentum is chosen so as to  have a regular solution to the field $x_3$; this kind of choice is also  used in \cite{kr}, \cite{st} (see the study of this configuration at finite temperature in \cite{mw}). However, in \cite{rm} a different  choice to momentum, which is nonzero,  yields a regular solution. We have detailed our choice in  Appendix D.

From this action it just follows that there exists a trivial solution to $y$, which is $y=0$. At IR, the fluctuated field, which we also denote as $y$, sees an $AdS_2$ spacetime with mass: $m^2=-\f{2B^2_2}{\rho^2+B^2_2(1+B^2_3)}$ as in the case without the Chern-Simon term in the action, which can violate the BF bound \cite{bf} when the following condition is met:
\be\label{cond_ir}
\f{\rho^2}{B^2_2}+B^2_3<7,
\ee
whereas the fluctuated field $y$ sees an $AdS_4$ at UV, but with mass, $m^2=-\f{2}{1+B^2_3-\Sigma^2/\alpha^2}$, which can go below the corresponding BF bound for the choice $B^2_3 < \f{\Sigma^2}{\alpha^2}-\f{1}{9}$. Since we do not want to have an instability at UV, this suggests putting a constraint on the magnitude of $B_3$ which is 
\be\label{cond_1}
B^2_3 > \f{\Sigma^2}{\alpha^2}-\f{1}{9}.
\ee

In order to compute the 1-pt function, we shall follow the same procedure as before and that of \cite{jkst}. Note that the function $f=(r^2+y^2)^{-2}$, and the Chern-Simon part of the action   at the core region becomes negligible and can be dropped  in eq(\ref{lin_dbi_cs}). So one essentially ends up with the  action   
\be
S_{core}=-\alpha\int  \sqrt{1+y'^2}\sqrt{7+\f{r^4}{(r^2+y^2)^2}},
\ee
where the field $y$ saturates the BF bound. The
 solution  admits the same $Log$ structure as before, which displays the scaling symmetry as well. At the far away one can find the solution to the linearized (around the trivial solution $ y=0$) equation of motion that follows eq(\ref{lin_dbi_cs}). The solution appears as

\be
f_{\pm}=u^{\f{1\pm i\alpha}{2}}{}_2F_1[\f{2-\beta\pm i\alpha}{8},\f{2+\beta\pm i\alpha}{8},1\pm\f{ i\alpha}{4},-u^4],
\ee
where 
\be\label{parameters_d3_d5_dbi_cs}
\alpha=\sqrt{\f{\rho^2_c-\rho^2}{\rho^2+B^2_2(1+B^2_3)}},~~~\beta=\sqrt
{\f{1+9(B^2_3-\Sigma^2/\alpha^2)}{1+(B^2_3-\Sigma^2/\alpha^2)}},~~~u=\delta ~r,~~~\delta=\bigg(\f{1+B^2_3-\Sigma^2/\alpha^2}{\rho^2+B^2_2(1+B^2_3)}\bigg)^{1/4},
\ee
 where ${}_2F_1[a,b,c,z]$ is the hypergeometric function and $\rho_c=B_2\sqrt{7-B^2_3}$. Since, we want the constants $\beta$ and $\delta$ to  be real implies a constraint on $B_3$ that is 

\be\label{cond_2}
B^2_3 >\Sigma^2/\alpha^2 -1/9.
\ee 
Recall $\Sigma/\alpha=\f{\mu_5 B_3}{T_5}=g_s B_3$, which means $B^2_3 >-\f{1}{9(1-g^2_s)}$. Note that $g_s$ is very small, using eq(\ref{cond_ir}) and eq(\ref{cond_2}) gives us the range of $B_3$ that is
\be
\Sigma^2/\alpha^2 -1/9< B^2_3 < 7-\f{\rho^2}{B^2_2}~~~\Rightarrow ~~~0 <B^2_3 < 7-\f{\rho^2}{B^2_2}.
\ee

If we compare eq(\ref{parameters_d3_d5_dbi_cs}) and eq(\ref{parameters_d3_d5}), then we find that the only difference to the linearized solution with and without the Chern-Simon action comes  at the structure to $\beta$ and $\delta$, from which it follows that the previous calculation of the 1-pt function goes through with just these modifications. But the condensate to exponential accuracy remains the same
\be\label{condensate_d3_d5_non_zero_rho_dbi_cs}
\sigma \sim e^{-\pi/\alpha}=e^{-\f{{\tilde c}}{\sqrt{\rho_c-\rho}}}=e^{-\f{c}{\sqrt{\nu_c-\nu}}},
\ee
where 
\be\label{condensate_d3_d5_non_zero_rho1_dbi_cs}
{\tilde c}=\pi\sqrt{\f{\rho^2+B^2_2(1+B^2_3)}{\rho_c+\rho}},~~~\nu=\f{\rho}{B_2\sqrt{1+B^2_3}},~~~c=\pi\sqrt{\f{\nu^2+1}{\nu+\nu_c}},~~~\nu_c=\f{\rho_c}{B_2\sqrt{1+B^2_3}}
\ee

\section{Broken phase: D5-D5}

The UV behavior of this configuration gives an $AdS_3$ space which saturates the BF bound for the field $y/r$ whereas at IR it sees an $AdS_2$, which can violate the BF bound when the $\f{\rho}{B_2} < \sqrt{3}$ condition is met. At the boundary the field $\Phi\equiv y/r$ can have an expansion of the form \cite{dss} and \cite{kos}
\be
\Phi=A/r+B/r~ Log~r
\ee
and we shall take $A$ as the vev of the operator dual to the bulk field $\Phi$ and $B$ as the source to which the operator couples to on the boundary.

The analysis for finding $A$ is precisely the same as is done for the D3-D5 case. In general, it is very difficult to solve non-linear equation of motion that follows from eq(\ref{action_b1_b2_b3}), so we shall try to find the solution at the core region and then find the solution in the far away region of the linearized equations of motion. In the overlapping region we shall find $A$ which is nothing but the condensate.

In the core region, when the scalar field saturates the BF bound,  the action is
\be
S=-\omega_1 T_5 B_2N_f\int \sqrt{1+y'^2}\sqrt{3+\f{r^2}{r^2+y^2}},
\ee
whose numerical solution with the boundary condition, $y(0)=0$, can be fitted to 
\be\label{log_sol}
y=\sqrt{r}[-a_0+a_1~log~r]=a_1
\sqrt{r}~Log~(r/r_0),~~~r_0=
e^{a_0/a_1},
\ee
for some constants  $a_0$ and $a_1$.
If we look at the linearized equation of motion at IR then it follows that the solution of it has precisely the structure of  eq(\ref{log_sol}) with the equation of motion showing the scaling symmetry as the action receives an overall multiple factor under scaling symmetry. 

The far away solution of the linearized equation of motion has the structure 
\be
f_{\pm}=u^{\f{1\pm i\alpha}{2}}{}_2F_1[\f{1\pm i\alpha}{4},\f{1\pm i\alpha}{4},1\pm\f{ i\alpha}{2},-u^2],
\ee
where 
\be
\alpha=\sqrt{\f{\rho^2_c-\rho^2}{\rho^2+B^2_2}},~~~u=\f{r}{\sqrt{\rho^2+B^2_2}}
\ee

Now proceeding as before, in order to fix the  boundary condition we set the bare quark mass to zero and the  asymptotic behavior of $f_n$ as 
\be
f_n\sim c_+f_++c_-f_-\rightarrow \sigma,~~~u\rightarrow \infty
\ee

The asymptotic forms of $f_{\pm}$ are
\be
f_+\rightarrow c_0+c_1 Log~u,~~~f_-\rightarrow {\tilde c}_0+{\tilde c}_1 Log~u,
\ee

which gives the necessary equations to fix the boundary conditions, 
\be
c_+=\-\f{\sigma {\tilde c}_1}{c_1{\tilde c}_0-c_0{\tilde c}_1},~~~c_-=\-\f{\sigma  c_1}{c_1{\tilde c}_0-c_0{\tilde c}_1}.
\ee

Using these ingredients, we find
\be
y=-f_n=\sqrt{u}\sigma\f{Y}{\alpha}~Sin\bigg(\f{\alpha}{2}Log~(u/u_1)\bigg),
\ee 
where $u_1=e^{2X/Y}$ and 
\bea
X&=&-\f{\Gamma[1/4]\Gamma[3/4](2\gamma+\pi+2\Psi(1/4))}{\Psi(\f{1} {4})^2+2\gamma(\pi+\Psi(\f{1} {4})-\Psi(\f{3} {4}))-\psi(\f{3} {4})^2
+\pi(\Psi(\f{1} {4})+\Psi(\f{3} {4}))+\Psi^{(1)}(\f{1} {4})+\Psi^{(1)}(\f{3} {4})}\nn 
Y&=&\f{4\Gamma[1/4]\Gamma[3/4]}{\Psi(\f{1} {4})^2+2\gamma(\pi+\Psi(\f{1} {4})-\Psi(\f{3} {4}))-\Psi(\f{3} {4})^2
+\pi(\Psi(\f{1} {4})+\Psi(\f{3} {4}))+\Psi^{(1)}(\f{1} {4})+\Psi^{(1)}(\f{3} {4})},\nn
\eea
where $\gamma$ is the Euler's constant, $\Psi(z)$ is the digamma function, and $\Psi^{(1)}(z)$ is the polygamma function of order one. Matching of the core solution and far away solution at $r=\xi$ gives the condition
\be
\sigma\sim\sqrt{\xi}\sim e^{-\pi/\alpha}=e^{-\f{{\tilde c}}{\sqrt{\rho_c-\rho}}}=e^{-\f{c}{\sqrt{\nu_c-\nu}}},
\ee
where 
\be
{\tilde c}=\pi\sqrt{\f{\rho^2+B^2_2}{\rho_c+\rho}},~~~\nu=\f{\rho}{B_2},~~~c=\pi\sqrt{\f{\nu^2+1}{\nu_c+\nu}}.
\ee

As suggested in \cite{jkst}, the Efimov states appear when we set the argument of trigonometric function $sin$ to a multiple of $\pi$, i.e., $n\pi$ and $n$ denotes the nth  states, and the condensate scales as $\sigma_n\sim e^{-\f{n\pi}{\alpha}}$.  

\section{Conclusion}

In this paper we have studied the quantum phase transition for  Dp-Dq brane configurations with external fields at zero temperature. We have chosen the intersecting 
brane configuration in such a way that it preserves some amount of supersymmetry in the absence of external fields and then we turned on the external electric and multiple magnetic fields on the world volume of the Dq brane. The result of analyzing the effective action of the Dq brane in the probe approximation using the DBI action  shows that some of the brane configurations can have a tachyonic scalar field. In fact, these scalar fields can also violate the BF bound at IR for some range of charge density and magnetic field. This sets in the instability of the system thereby forcing the system to undergo a phase transition. The interesting brane configurations  considered are D3-D5, D5-D5, and D3-D7 with electric and magnetic fields.

The outcome of this study is that one can have a BKT-type phase transition with the D3-D5 brane configuration even in the absence of any charge density but only with multiple magnetic fields. In \cite{jkst}, it was shown that, for the BKT type of transition to happen, the charge density and magnetic fields are essential. Here we have generalized that and have shown even with  multiple magnetic fields alone is enough to make the transition to occur. We have found yet another example of    brane configuration --that of the D5-D5 system which 
also exhibits the BKT type of transition.

 It is worth  emphasizing that the Dp-Dq brane configuration that we started out with before turning on the external electric and magnetic fields preserves some amount of supersymmetry. But after turning on the external fields the system breaks the supersymmetry and that is why the instability comes into picture. It is certainly interesting to study the situations where  the external electric and magnetic fields are turned on the Dp-Dq brane configuration which are nonsupersymmetric to start with.

The D3-D7 brane configuration with zero density but with two constants magnetic fields of strength $B_2$ and $B_3$ also has a tachyonic scalar field in the linearized approximation to the DBI action. The mass of the scalar field does in fact go below the  BF bound for any real valued choice of the magnetic fields and prompting a phase transition to occur.
We have calculated the condensate and found it's dependence on the magnetic field $B_2$, which says that the condensate vanishes for zero value to $B_2$.

 It is suggested in \cite{jkt} that the transition for the D3-D7 system could possibly be a second order  but not higher than third order (suggested to be third order in the absence of magnetic field \cite{fl}) and is described by mean field exponents, which is very interesting to study further along those lines,  so also to find  the complete  phase diagram both at zero and nonzero temperature and chemical potential plane as initiated in \cite{egkm}. \\

\section{Acknowledgment}
It is a pleasure to thank Ofer Aharony and Andreas Karch  for several suggestions, clarifications and for  
going through the manuscript. Thanks  to the anonymous referee for several useful clarifications and suggestions. Thanks to Bum-Hoon Lee for two extended discussions on the material presented. Also to K. Jensen for a suggestion and the members of  CQUeST for  their help.

This work was supported by the Korea Science and Engineering Foundation (KOSEF) grant funded by the Korea
government (MEST) through the Center for Quantum Spacetime (CQUeST) of Sogang University with Grant No, R11-2005-021.

{\bf Note added}-- We learned  of the work of K. Jensen \cite{kj}, which also studies the BKT transition and with which there are some overlaps.

\section{Appendix: A simple derivation of BKT type scaling}

In this Appendix we provide a simple sketch of seeing  BKT-type behavior in $1/r^{2}$ potential. The main idea behind such a derivation lies in the fact that a scalar field in $AdS$ space looks precisely the same as that of the Schroedinger equation with $v_0/r^2$ potential, where $v_0$ is a constant.

The Schroedinger equation in $d$ dimension, 
\be
\f{d^2}{dr^2}\psi(r)+\f{(d-1)}{r}\f{d}{dr}\psi-\f{v_0}{r^2} \psi(r)=-E \psi(r)
\ee
has the scaling symmetry $r\rightarrow \Lambda ~r$ and $E\rightarrow \Lambda^{-2}~E$. 
The solution with two constants $c_1,~c_2$
\be
\psi=r^{1-d/2}[c_1 J_{\sqrt{(1-d/2)^2+v_0}}(\sqrt{E}r)+c_2 Y_{\sqrt{(1-d/2)^2+v_0}}(\sqrt{E}r)]
\ee
where the functions $J_{\mu}(z),~Y_{\mu}(z)$ are   Bessel functions of the first and second kind respectively.
For $ v_0 \geq -(1-d/2)^2$, the solution  is real and for $ v_0 < -(1-d/2)^2$ it becomes complex and there is no bound on the Hamiltonian \cite{gr}. Hence, it becomes unstable. 

Let us look at the equation of motion of a  minimally coupled scalar field, $\phi$, with mass $m$ in $AdS_{D+1}$
\be
\phi''+\f{D+1}{r}\phi'-\f{m^2}{r^2}\phi=0,
\ee

Comparing the Schroedinger equation in $v_0/r^2$ potential with that of the equation of motion of the minimally coupled  scalar field gives the condition for stable solution 
\be
v_0 \geq -(1-d/2)^2~~~ \Longrightarrow ~~~m^2 \geq -\f{D^2}{4},
\ee
which is nothing but the BF bound. Similarly, the instability arises, when this
\be
v_0 <-(1-d/2)^2~~~ \Longrightarrow ~~~m^2 < -\f{D^2}{4}
\ee
condition is satisfied. If we set  $v_0$ to go below $-(1-d/2)^2$, then the level of the Bessel function becomes imaginary, which results in  oscillatory solution.

Using a field redefinition $\psi=\xi r^{-\f{(d-1)}{2}}$,  we can rewrite the Schroedinger equation with zero energy as
\be
\xi''-\f{\beta}{r^2}\xi=0,
\ee
where $\beta=(1-d/2)^2+v_0-1/4$. Let us look at  the situation  $v_0 <-(1-d/2)^2$ and  denote 
$v^{BF}_0=-(1-d/2)^2$,  which means $\beta=v_0-v^{BF}_0-1/4$ and  the solution with the boundary condition, $\xi(r_{0})=0$, implies
\be
\xi=\sqrt{r}~ sin[\sqrt{v^{BF}_0-v_0}~Log(r/r_{0})].
\ee

The reason for getting a real solution even for $v_0 <-(1-d/2)^2$ is that the space is now truncated to  
$ r_{\infty}\leq~r~\leq r_{0}$. If we demand that  the field $\xi$ satisfies  the Dirichlet boundary condition at $r_{\infty}$, then it gives 
\be
\f{1}{r^2_{\infty}}=\f{1}{r^2_{0}}~e^{-\f{2\pi}{\sqrt{v^{BF}_0-v_0}}},
\ee
which is obtained by fitting the solution $\xi$ to  half the period of trigonometric function.  The purpose of putting the Dirichlet boundary condition at UV, $r=r_{0}$, is  that the potential $v_0/r^2$ diverges as one approaches the origin, which needs to be regularized \cite{klss}. The way we do that is by putting a hard wall cutoff at, $r=r_0$, thereby truncating the space. From the scaling symmetry of the Schroedinger equation, it just follows that   the inverse square of the radial coordinate, $1/r^2$ plays the role of energy scale, which means we can interpret $1/r^2_0$ as the energy scale at UV and $1/r^2_{\infty}$ as the scale at IR.  So fixing the UV boundary condition naturally generated an IR scale \cite{klss} and \cite{ilms}.

There is another way to see the appearance of BKT scaling. Let us look at the solution for which   $v_0 <-(1-d/2)^2$, which means the label of the Bessel function is imaginary, and choose the solution as 
\be
\psi(r)=r^{1-d/2}[c_1 J_{i\sqrt{|\alpha|}}(\sqrt{E}r)+c_2 J_{-i\sqrt{|\alpha|}}(\sqrt{E}r)],
\ee 
where $\alpha=(1-d/2)^2+v_0$, Now taking the coefficient $c_1$ and $c_2$ as equal in magnitude but opposite in sign, i.e., $c_1=-c_2$, gives the solution near to the origin as
\be
\psi(r)=2i~c_1~r^{1-d/2}~sin[|\alpha|~log~\sqrt{E}r].
\ee
Let us truncate the space from $0\leq r\leq \infty$ to $r_0\leq r \leq r_{\infty} $ and impose the boundary condition that $\psi(r_0)=0$. This results in
\be
E=\f{1}{r^2_0}e^{\f{-2\pi}{|\alpha|}}=\f{1}{r^2_0} ~e^{-\f{2\pi}{\sqrt{v^{BF}_0-v_0}}},
\ee

The radial coordinate $r$ cannot also extend all the way to infinity as there is a finite periodicity associated to the trigonometric function. 

\section{Appendix B: Potential energy}

In this Appendix, we shall calculate the potential energy seen by the probe Dq brane in the Dp-Dq brane configuration.
Generically, it is very difficult to calculate it analytically, because  finding an analytic solution to the equation of motion that follows from DBI action is very difficult. So, the approach that we shall adopt is to find it only after linearizing the nonlinear DBI action around the trivial solution, $y=0$. 

\subsection{D3-D5 system}
First we shall calculate the potential energy for the D3-D5 system without the Chern-Simon term, in which case 
the linearized equation of motion that follows from eq(\ref{quadratic_action_b1_b2_b3}) for nonzero charge density and magnetic fields 
\be
r^2[\rho^2+(r^4+B^2_2)(1+B^2_3)]y''(r)+2 r^5 (1+B^2_3)y'(r)+2(B^2_2-B^2_3 r^4) y(r)=0.
\ee
By doing a field redefinition $y=r\chi(r)$, we can bring the equation of motion to 
\be
\chi''+\bigg(\f{p'(r)}{p(r)}+\f{2}{r}\bigg)\chi'+\bigg(\f{p'(r)}{rp(r)}-\f{q(r)}{p(r)}\bigg) \chi=0,
\ee
where 
\be
\f{p'(r)}{p(r)}=\f{2r^3(1+B^2_3)}{[\rho^2+(r^4+B^2_2)(1+B^2_3)]},~~~ -\f{q(r)}{p(r)}=\f{2(B^2_2-B^2_3 r^4)}{r^2[\rho^2+(r^4+B^2_2)(1+B^2_3)]}
\ee 

We can recast the equation of motion to $\chi$ in the Schroedinger equation form $\f{d^2\chi(u)}{du^2}+[\lambda-Q(r)]\chi(u)=0$, with the eigenvalue
$\lambda$. By defining a new coordinate system $u=\int \f{dr}{r^2p(r)}$,  the potential energy for zero eigenvalue
\be\label{potential_energy}
Q(r)=-r^3 p (p'-r q)=-2r^2(B^2_2+r^4)
\ee

Let us  include the effect of the Chern-Simon term, eq(\ref{csterm_d3_d5}), and the inclusion of it changes the structure to action, which is governed by eq(\ref{lin_dbi_cs}). The resulting linearized equation of motion that follows around the trivial solution to $y$ yields
\be
r^2\bigg[\rho^2+(r^4+B^2_2)(1+B^2_3)-r^4\f{\Sigma^2}{\alpha^2}\bigg]y''(r)+2 r^5 \bigg(1+B^2_3-\f{\Sigma^2}{\alpha^2}\bigg)y'(r)+2\bigg[B^2_2-\bigg(B^2_3-\f{\Sigma^2}{\alpha^2}\bigg) r^4\bigg] y(r)=0.
\ee 

Proceeding as previously, we find  the field $\chi(r)=\f{y(r)}{r}$ obeys the Schroedinger equation $\f{d^2\chi(u)}{du^2}-Q(r)\chi(u)=0$, with zero eigenvalue in the coordinate $u=\int \f{dr}{r^2p(r)}$, 
where the function 
\be
p(r)=\sqrt{\rho^2+(r^4+B^2_2)(1+B^2_3)-r^4\f{\Sigma^2}{\alpha^2}},
\ee

and the potential energy
\be\label{potential_energy_cs}
Q(r)=-r^3 p (p'-r q)=-2r^2(B^2_2+r^4)
\ee

Even though it looks like the forms of  potential energy eq(\ref{potential_energy}) and eq(\ref{potential_energy_cs})
are same, they are not. Recall that the coordinate $r$ is related to $u$ differently in these two cases. The difference is due to the form of $p(r)$.

Let us look at the potential energy eq(\ref{potential_energy_cs}) at IR, which goes as 
$Q(r)\sim -2 r^2 B^2_2$. The expression to $u$ at IR 
is $u\simeq -\f{1}{\sqrt{\rho^2+B^2_2}}\times \f{1}{r}$, which says that the potential  energy at IR is 
$Q(u)\simeq - \bigg(\f{B^2_2}{\rho^2+B^2_2}\bigg) \bigg(\f{2}{u^2}\bigg)$. Similarly at UV the potential energy behaves as $Q(u)\simeq \f{1}{1+B^2_3-\Sigma^2/\alpha^2} \times \f{2}{9u^2}.$

\subsection{D5-D5 system}
Once again the linearized equation of motion that follows from eq(\ref{quadratic_action_b1_b2_b3}) for nonzero charge density and magnetic fields 
\be
r^2[\rho^2+r^2+B^2_2]y''(r)+ r^3 y'(r)+B^2_2 y(r)=0.
\ee

After a field redefinition, $y=r \chi(r)$, the equation becomes
\be
\chi''(r)+\bigg(\f{r}{[\rho^2+r^2+B^2_2]}+\f{2}{r}\bigg)\chi'(r)+\f{(r^2+B^2_2)}{r^2[\rho^2+r^2+B^2_2]}\chi(r)=0,
\ee
and coordinate transformation, $u=\int \f{dr}{r^2\sqrt{\rho^2+r^2+B^2_2}}=-\f{\sqrt{\rho^2+r^2+B^2_2}}{r(\rho^2+B^2_2)}$, brings the equation of motion to the following form:
\be
\f{d^2\chi(u)}{du^2}-Q(r) \chi(u)=0, 
\ee 
where the potential 
\be
Q(r)=-r^3 p (p'-r q)=-r^2(B^2_2+r^2),
\ee
has a power law behavior. Now, using the relation between $r$ and $u$ i.e. $r^2=\f{\rho^2+B^2_2}{u^2(\rho^2+B^2_2)^2-1}$, gives us the potential as
\be
Q(u)=\f{(\rho^2+B^2_2)}{[u^2(\rho^2+B^2_2)^2-1]^2}\times \bigg[B^2_2\bigg(u^2(\rho^2+B^2_2)^2-1\bigg)+\rho^2+B^2_2\bigg].
\ee

\subsection{D3-D7 system}

Here we shall be evaluating the potential energy only in the zero charge density case but with nonzero magnetic fields. The linearized equation of motion that follows from eq(\ref{quadratic_action_b1_b2_b3}) is
\be
r^2(B^2+r^4)(1+B^2_3)y''(r)+r(B^2+3r^4)(1+B^2_3)y'(r)+2(B^2_2-B^2_3 r^4)y(r)=0
\ee

Doing a field redefinition, $y(r)=r\chi(r)$,  brings down the equation to
\be
\chi''+\bigg(\f{p'(r)}{p(r)}+\f{2}{r}\bigg)\chi'+\bigg(\f{p'(r)}{rp(r)}-\f{q(r)}{p(r)}\bigg) \chi=0,
\ee
where 
\be
\f{p'(r)}{p(r)}=\f{B^2+3r^4}{r(B^2+r^4)},~~~
-\f{q(r)}{p(r)}=\f{2(B^2_2-B^2_3 r^4)}{r^2(B^2+r^4)(1+B^2_3)}.
\ee
Defining a new coordinate $u=\int \f{dr}{r^2p(r)}=-\f{\sqrt{B^2_2+r^4}}{2B^2_2r^2}$, gives us $r^4=\f{B^2_2}{4B^2_2u^2-1}$, which brings the equation to the Schroedinger equation form, $\f{d^2\chi(u)}{du^2}-Q(r) \chi(u)=0$, with the  potential 
\be
Q(r)=-\bigg(\f{3+B^2_3}{1+B^2_3}\bigg)r^4(B^2_2+r^4),
\ee
which in the $u$ coordinate system can be written as 
\be
Q(u)=-4B^8_2\bigg(\f{3+B^2_3}{1+B^2_3}\bigg)\bigg(\f{u}{(1-4B^4_2u^2)}\bigg)^2.
\ee

\section{Appendix C: Chern-Simon term}

In this Appendix, we shall calculate the contribution of the Chern-Simon part of the action to some of the probe brane action.
The Chern-Simon action to the Dq brane  \cite{mrd}
\be
\mu_q \int \sum_n \bigg[C_{(n)}\w e^B\bigg]\w e^F,
\ee
where $\mu_q$ is the charge of the D1 brane and [~] denotes pullback of the background bulk fields onto the world volume of the  Dq brane. As  suggested in \cite{bds} it is the $B+F$ term that is gauge invariant not just the gauge field strength $F$, which means turning on $F$ implies turning on $B$ as well. The configurations that we shall deal with here are D3-D5, D3-D7, and D5-D5.\\

D3-D5 brane configuration:\\

In this case the only nonzero and nontrivial RR potential comes from the 4-form potential $C_4$, whose structure is
\be
C_4=\f{1}{f(r,y)}dt\w dx_1\w dx_2\w dx_3,
\ee
where $f(r,y)$ is written in eq(\ref{dp_brane}) and with our choice to $F_2$ as in eq(\ref{2_b_field_A}) results in the contribution  as
\be\label{c_4}
\mu_5 \int  \bigg[C_{(4)} \w B\bigg]\w F
\ee

Let us recall that the D5 brane world volume coordinates are  along $[t,~x_1,~x_2,~r,~\theta,~\phi]$. Evaluating eq(\ref{c_4}) it in the static gauge choice and exciting only the scalar field $y(r)$, it just follows that at this order in $\alpha'$ the Chern-Simon part of the action does not contributes, where we have assumed the linear embedding $x_3$ is constant, but this is not a consistent approximation. 

Once we excite $x_3(r)$ along with $y(r)$, then the contribution to $C_4$ and the Chern-Simon action 
\bea\label{csterm_d3_d5}
&&C_4=\f{x'_3(r)}{f(r,y)}dt\w dx_1\w dx_2\w dr,\nn
&&\mu_54\pi N_f B_3\int\f{x'_3(r)}{f(r,y)}dt\w dx_1\w dx_2\w dr
\eea

Because of this flux, which is proportional to  $B_3$,  the induced metric on the D5 brane is
\be\label{ind_metric_d5}
ds^2(ind)=f^{-1/2}\bigg[-dt^2+dx^2_1+dx^2_2+\bigg(x'^2_3(r)+f(1+y'^2)\bigg)dr^2\bigg]+
f^{1/2}r^2d\Omega^2_2
\ee 

D3-D7 brane configuration:\\

In this case the only term that can contribute is 
\be\label{c_4_d3_d7}
\f{\mu_5}{2} \int  [C_{(4)}] \w ([B]+ F)^2
\ee

The world volume coordinates for the D7 brane are $[t,~x_1,~x_2, ~x_3,~r,~\theta,~\phi,~\psi]$, where the induced metric on the D7 brane is
\be
ds^2=f^{-\f{1}{2}}[-dt^2+dx^2_1+dx^2_2+dx^2_3]+f^{\f{1}{2}}[dr^2(1+y'(r)^2)+r^2(d\theta^2+s^2_{\theta}d\phi^2+
s^2_{\theta}s^2_{\phi}d\psi^2)]
\ee

Now using the structure to $F_2$ (with $B_3=0$) from eq(\ref{2_b_field_A}), results  that, at the leading order in $\alpha'$, there is not any contribution coming from the Chern-Simon action.\\

D5-D5 brane configuration:\\

In this case the only nonzero RR potential is the 6-form, $C_6$ form potential
\be
C_6=\f{1}{f(r,y)}dt\w dx_1\w dx_2\w dx_3\w dx_4\w dx_6
\ee 
and it's magnetically dual 2-form defined as~ ${\star _{10}}dC_6=e^{-\Phi}dC_2$, with
\be
dC_2=\f{2}{3}ry\bigg(\p_r f^{3/2}dy-\p_y f^{3/2}dr \bigg)\w d\theta\w d\phi
\ee
So the relevant terms are in the action are
\be\label{cs_action_d5-d5}
\mu_5\int \bigg(\f{1}{2}[C_2]\w ([B]+F)^2+ [C_6]\bigg)
\ee

We recall now that the world volume coordinates of the probe brane are $[t,~x_1,~x_2,~x_3,~r,~\theta]$. With  the static gauge choice and exciting  the scalar field $y(r)$, this implies that the last term in eq(\ref{cs_action_d5-d5}) is not going to contribute. Using the gauge field as written in eq(\ref{1_b_field}) says that the first term in eq(\ref{cs_action_d5-d5}) too is not  going to contribute. So, at the leading order in $\alpha'$ there is not any nontrivial contribution coming from the Chern-Simon action.\\

D2-D4,~D4-D6~ and D5-D7 brane configurations:\\

These brane configurations are T dual to D3-D5 brane configuration modulo the number of the constant magnetic fields that are turned on. In all these cases the flavor brane is wrapped on a two sphere perpendicular to the color brane, which means we can turn on a flux on this two sphere, which is proportional to $B_3$. The only big difference among these brane configurations is they live in 1+1, 3+1 and 4+1 dimensional spacetime for D2-D4,~D4-D6,~ and D5-D7 brane configurations, respectively.

Now turning on the flux on $S^2$ suggests there should be a nontrivial term that contributes to the Chern-Simon part of the action which essentially means we have to excite a scalar field that is parallel to the color brane but perpendicular to the field theory that we are interested in.

So the Chern-Simon term for the flavor Dq brane, for this kind of brane configuration, is
\be
\Sigma_q \int \f{x'_{q-2}(r)}{f(r,y)}dt\w dx_1\w \cdots\w dx_{q-3}\w dr,
\ee

where $\Sigma_q=\mu_q 4\pi N_f B_3$. In writing down this equation we have already done the integration over the two sphere.

The inclusion of the Chern-Simon part of the action to the total action for this kind of brane configurations does not change the results stated in eq(\ref{table_zero_density}) and eq(\ref{table_density}). The way to see it is as follows: The total action for these cases is 
\be
S=-\alpha\int \sqrt{x'^2f^{-1}+1+y'^2-A'^2_0}\sqrt{(1+fB^2_1)(f^{-1}+B^2_2)(fr^4+B^2_3)}+\Sigma\int \f{x'}{f},
\ee
where we have rewritten $x_{q-2}$ as $x$ and kept all  three different kinds of magnetic fields to describe all the three different kinds of brane configurations with one action. So one has to keep in mind that for D2-D4 configuration $B_1$ and $B_2$ are zero and for D4-D6 $B_1$ is zero whereas for D5-D7 all $B_i$'s are nonzero.

From this action it follows trivially that the momenta associated to $x$ and $A_0$ are constants. So after doing the Legendre transformation $S_L=S-\int \f{\delta S}{\delta A'_0}A'_0-\int  \f{\delta S}{\delta x'_{q-2}}x'_{q-2}$, we ended up with the action for zero momentum to $x$ as
\bea
S^{D2-D4}_L&=&-\alpha\int \sqrt{1+y'^2}\sqrt{\rho^2+fr^4+B^2_3-f^{-1}\f{\Sigma^2}{\alpha^2}},\nn
S^{D4-D6}_L&=&-\alpha\int \sqrt{1+y'^2}\sqrt{\rho^2+(f^{-1}+B^2_2)(fr^4+B^2_3)-f^{-1}\f{\Sigma^2}{\alpha^2}}, \nn
S^{D5-D7}_L&=&-\alpha\int \sqrt{1+y'^2}\sqrt{\rho^2+(1+fB^2_1)(f^{-1}+B^2_2)(fr^4+B^2_3)-f^{-1}\f{\Sigma^2}{\alpha^2}}.\nn
\eea

Now the equation of motion that follows from the linearized fluctuation  to $y$ around its trivial solution does not change the result to the equation of motion that one obtains without the Chern-Simon term. The simplest way to see is to use  $f=(r^2+y^2)^{\f{p-7}{2}}$ and use the fact that we are using color branes for which $p < 7$ in the last term of the second square root. Essentially, this term gives 
subdominant contribution, so one can safely drop these terms and find the equation of motion and hence the mass square to the fluctuated field. The result is as stated in 
 eq(\ref{table_zero_density}) and eq(\ref{table_density}).

For completeness, the solution to $x_{q-2}=x$ and $A_0$ can be found: 
\bea
A'^2_0&=&\f{\alpha^2\rho^2(1+y'^2)}{\alpha^2(1+fB^2_1)(f^{-1}+B^2_2)(fr^4+B^2_3)+\rho^2\alpha^2-f^{-1}(C_{q-2}f-\Sigma)^2},\\
x'^2_{q-2}&=&\f{(C_{q-2}f-\Sigma)^2(1+y'^2)}{\alpha^2(1+fB^2_1)(f^{-1}+B^2_2)(fr^4+B^2_3)+\rho^2\alpha^2-f^{-1}(C_{q-2}f-\Sigma)^2},
\eea

by solving these equations with  keeping in mind that,  for the D2-D4 case, the only nonvanishing field is $B_3$,  for D4-D6 the nonvanishing fields are $B_2$ and $B_3$, and for D5-D7 case all of the magnetic fields are nonzero. It is interesting to note  the choice for the vanishing momentum associated to $x_{q-2}$ i.e. $C_{q-2}=0$ yields a regular solution to $x_{q-2}$.

\section{Appendix D: Solution to D3-D5 with a scalar field}

The induced metric on D5 brane is
\be\label{ind_metric_d5}
ds^2(ind)=f^{-1/2}\bigg[-dt^2+dx^2_1+dx^2_2+\bigg(x'^2_3(r)+f(1+y'^2)\bigg)dr^2\bigg]+
f^{1/2}r^2d\Omega^2_2
\ee 

Using the trivial solution for $y$ that is $y=0$ with $C=0$ and the expression to $x'_3$ from eq(\ref{sol-x3-A0}) results in

\bea
x'^2_3&=&\f{\Sigma^2}
{\alpha^2[\rho^2+B^2_2(1+B^2_3)]+r^4[\alpha^2(1+B^2_3)-
\Sigma^2]},\nn 
x'^2_3+f&=&\f{\alpha^2}{r^4}\f{[\rho^2+B^2_2(1+B^2_3)+r^4(1+B^2_3)]}{[\alpha^2(\rho^2+B^2_2(1+B^2_3))-r^4(\Sigma^2-\alpha^2(1+B^2_3))]}
\eea  

From this, one can check that indeed $C=0$ results in a regular solution both at IR and UV. The exact solution  is in the form of an incomplete elliptic integral.
The quantity $x'^2_3+f$ behaves at IR and UV as
\bea
x'^2_3+f &\longrightarrow& \f{1}{r^4},~~~r\rightarrow 0,\nn
&\longrightarrow& \f{1}{r^4}\times \Bigg(\f{\alpha^2(1+B^2_3)}{\alpha^2(1+B^2_3)-\Sigma^2}\Bigg)\equiv \f{X}{r^4},~~~r\rightarrow \infty
\eea

So it follows that at IR the induced metric on the D5 brane is 
$AdS_4\times S^2$, where the size of both  $AdS_4$ and $S^2$ are unity. Recall that we are working in units for which $R=1$. At UV, even though the spacetime has the same topology as at IR but the sizes of $AdS_4$ and $S^2$ are different. In fact the size of $AdS_4=\sqrt{X}$ whereas the radius of $S^2$ is still unity. Note that the induced metric somewhere in the interior is not $AdS$ and hence  {\em a priori} the metric do not preserves conformal symmetry over the  entire spacetime. Moreover, one can very easily notice that the dimension of density $\rho$ and the magnetic field that is turned on along the field theory direction that is $B_2$ are the same. So even in the absence of  density the $B_2$ field sets a  scale and, hence, there occurs a breakdown of the conformal symmetry. Roughly, one can think that turning on a constant magnetic field along the field theory direction induces noncommutativity among the spatial coordinates of the theory and  sets a scale which is proportional to $B_2$ \cite{sw}.

Let us see the mass formula for the fluctuation to $y$ at IR, which sees an $AdS_2$ spacetime,
\be
m^2=-\f{2B^2_2}{\rho^2+B^2_2(1+B^2_3)},
\ee
with the size of $AdS$ set to unity as we saw in the previous section. For zero density this mass formula reduces to $m^2=-\f{2}{1+B^2_3}$. Now if we want to impose the condition that this fluctuation should violate the BF bound means $m^2 <-\f{1}{4}$, with the same size to $AdS_2$, which essentially set the condition $B^2_3 < 7$.

Let us see what happens if both $\rho=0=B_2$, as studied in \cite{rm}. In this case the quantity
\be
x'^2_3+f=\f{X}{r^4},
\ee

Hence, the induced metric is  $AdS_4\times S^2$, confirming the result of \cite{rm} even though it is calculated there with a nonzero momentum to $x_3$, and the conformal symmetry is not broken by the magnetic field $B_3$, as it is dimensionless in our units. Moreover, it just rescales the size of $AdS_4$.

\end{document}